\documentclass[fleqn,usenatbib]{mnras}

\usepackage{newtxtext,newtxmath}

\usepackage[T1]{fontenc}

\DeclareRobustCommand{\VAN}[3]{#2}
\let\VANthebibliography\thebibliography
\def\thebibliography{\DeclareRobustCommand{\VAN}[3]{##3}\VANthebibliography}

\usepackage{graphicx}	
\usepackage{amsmath}	

\usepackage{booktabs}
\usepackage{rotating}
\usepackage{caption}
\usepackage{subcaption}
\usepackage{hyperref}

\title[Recovering Luminosity and Radii of Host Stars]{Testing Multiband ($G$, $G_{\rm BP}$, $G_{\rm RP}$, $B$, $V$ and $TESS$) Standard Bolometric Corrections by Recovering Luminosity and Radii of 341 Host Stars}

\author[Eker and Bak{\i}\c{s}]{
Eker, Z.\thanks{E-mail: eker@akdeniz.edu.tr (Z. Eker)}, 
Bak{\i}\c{s}, V.\thanks{E-mail: volkanbakis@akdeniz.edu.tr (V. Bak{\i}\c{s})}
\\
Department of Space Sciences and Technologies, Faculty of Sciences, Akdeniz University, 07058, Antalya, Turkiye\\
}

\date{Accepted XXX. Received YYY; in original form ZZZ}

\pubyear{2023}

\begin{document}
\label{firstpage}
\pagerange{\pageref{firstpage}--\pageref{lastpage}}
\maketitle

\begin{abstract}
Main-sequence bolometric corrections ($BC$) and a standard $BC-T_{\rm eff}$ relation are produced for TESS wavelengths using published physical parameters and light ratios from SED models of 209 detached double-lined eclipsing binaries. This and previous five-band (Johnson $B$, $V$, Gaia $G$, $G_{\rm BP}$, $G_{\rm RP}$) standard $BC - T_{\rm eff}$ relations are tested by recovering luminosity ($L$) of the most accurate 341 single host stars (281 MS, 40 subgiants, 19 giants and one PMS). Recovered $L$ of photometry are compared to $L$ from published $R$ and $T_{\rm eff}$. A very high correlation ($R^2$ = 0.9983) is achieved for this mixed sample. Error histograms of recovered and calculated $L$ show peaks at $\sim$2 and $\sim$4 per cent respectively. The recovered L and the published $T_{\rm eff}$ were then used in $L = 4\pi R^2 \sigma T^4_{\rm eff}$ to predict the standard $R$ of the host stars. Comparison between the predicted and published R of all luminosity classes are found successful with a negligible offset associated with the giants and subgiants. The peak of the predicted $R$ errors is found at 2 per cent, which is equivalent to the peak of the published $R$ errors. Thus, a main-sequence $BC-T_{\rm eff}$ relation could be used in predicting both $L$ and $R$ of a single star at any luminosity class, but this does not mean $BC-T_{\rm eff}$ relations of all luminosity classes are the same because luminosity information could be more constrained by star’s apparent magnitude $\xi$ than its $BC$ since $m_{\rm Bol}=\xi + BC_\xi$.
\end{abstract}




\section{Introduction}

Luminosity ($L$) is not an observable parameter for a star. There are only one direct and two indirect methods of obtaining $L$. The direct method uses the Stefan-Boltzmann law ($L = 4 \pi R^2 \sigma T_{\rm eff}^4$) for calculating a luminosity directly from a given radius ($R$) and effective temperature ($T_{\rm eff}$) of a star. Therefore, the most accurate stellar $L$ comes from observationally determined most accurate $R$ and $T_{\rm eff}$ values which would be available from Detached Double-lined Eclipsing Binaries \citep[DDEBs;][]{Andersen1991, Torres2010b, Eker2014, Eker2015, Eker2018}. The typical accuracy of a calculated $L$ is 8.2–12.2 per cent \citep{Eker2021b}.

The first of the two indirect methods requires mass and pre-determined classical mass-luminosity relation (MLR) in the form of $L \propto M^\alpha$, where the typical accuracy is 17.5-37.99 per cent \citep{Eker2018,Eker2021b}. In early times, especially after the discovery of main-sequence MLR independently by \cite{Hertzprung1923} and \cite{Russel1923}, the MLR was claimed to be one of the most prominent empirical laws of nature by \cite{Eddington1926} and \cite{Gobovits1938} and it was used either predicting $L$ from $M$ or $M$ from $L$ at least until the middle of the 20th century or perhaps until \cite{Andersen1991} was objecting it. Those were the times the observational accuracy of $L$ (or $M$) was not high enough to distinguish the true $L$ (or $M$) from the average  $L$ (or $M$) for a given $M$ (or $L$) of a main-sequence star. 

The second of the two indirect methods requires an absolute visual magnitude and a pre-estimated bolometric correction ($BC$) to compute the absolute bolometric magnitude of a star in the first step as $M_{\rm Bol}$ = $M_{\rm V}$ + $BC_{\rm V}$ and then to obtain its $L$ in SI units from

\begin{equation}
    M_{\rm Bol} = -2.5 \times log L + 71.197 425 ...,
	\label{eq:mbol1}
\end{equation}
where the typical accuracy of $L$ is about to 10-13  per cent \citep{Eker2021b,Eker2022,Bakis2022} which is equivalent to the uncertainty of $BC_{\rm V}$ if  the uncertainty of the  absolute visual magnitude is negligible and the $BC_{\rm V}$ is a standard $BC_{\rm V}$. Otherwise, that is if the $BC_{\rm V}$ is not a standard $BC_{\rm V}$, an additional uncertainty 10 per cent or more \citep{Eker2021b,Eker2022} would arise due to the complexities caused by the three paradigms (the $BC$ scale is arbitrary, $BC$ values must always be negative, and the bolometric magnitude of a star ought to be brighter than its $V$ magnitude) which are not valid since 2015 \citep{Eker2022}. It was \cite{Torres2010a} who first noticed inconsistencies due to improper usage of $M_{Bol,\odot}$ and tabulated $BC_{\rm V}$ values that may lead to errors of up to 10 per cent or more in the derived $L$ equivalent to about 0.1 mag or more in the bolometric magnitudes.

Disagreed bolometric corrections ($BC_{\rm V}$), which are primarily in tabulated form, had been used for about a century \citep{Kuiper1938,Popper1959,Johnson1966,Flower1996,Bessel1998,Girardi2008,Sung2013}. International Astronomical Union was aware of the problem, thus, issued a general assembly resolution\footnote{\url{https://www.iau.org/static/resolutions/IAU2015_English.pdf}}  hereafter IAU 2015 GAR B2, to solve the problems associated with the arbitrariness attributed to the zero point constants of the $M_{\rm Bol}$ and $BC$ scales. This revolutionary document \citep{Eker2022} appears to be ignored or its full potential was not understood properly since some authors continued the old tradition \citep{Casagrande2018,Andrae2018,Chen2019,Eker2020} under the influence of the paradigms. The revolution has been noticed first by \cite{Eker2021a} who revised the definition of standard bolometric correction according classical definition $BC_{\rm V} = M_{\rm Bol} - M_{\rm V}$  where $M_{\rm V}$ is calculated from an apparent magnitude, parallax, and interstellar extinction while $M_{\rm Bol}$ is calculated according to IAU 2015 GAR B2 (Eq.\ref{eq:mbol1}) or using
\begin{equation}
    M_{\rm Bol} = M_{bol,\odot}-2.5 \times log L/L_\odot
	\label{eq:mbol2}
\end{equation}
in which the nominal\footnote{Actual values are $M_{Bol,\odot} = 4.739 997 ...$ mag and $L_\odot = 3.8275(\pm0.0014)\times10^{26}$ W} solar values $M_{Bol,\odot} =$ 4.74  mag and $L_\odot=$ 3.828$\times10^{26}$ W should be used during pre-computation of a $BC$  because the zero point constant of the bolometric magnitude scale $C_{\rm Bol} = M_{Bol,\odot} + 2.5 log L_\odot = 71.197 425 ...$ has been fixed by the General Assembly of IAU in 2015 meeting.

A predicted $L$ according to the third method using an absolute magnitude and a standard $BC$ could be called standard $L$ while a computed $L$ according to the  Stefan-Boltzmann law is standard by the definition. Investigating typical and limiting accuracies of the three methods of obtaining stellar $L$ in the era after Gaia, \cite{Eker2021b} claimed that it is possible to predict a standard $L$ within a few per cent if the pre-required $BC_{\rm V}$ is measured directly from a high signal to noise ratio spectrum according to the following definition of $BC_{\rm V}$

\begin{equation}
    BC_{\rm V}=2.5log\frac{f_{\rm V}}{f_{\rm Bol}}+C_2=2.5log\frac{\int_{0}^{\infty} S_\lambda(V)f_\lambda d\lambda}{\int_{0}^{\infty} f_\lambda d\lambda}+C_2,
	\label{eq:bc}
\end{equation}
where $S_\lambda(V)$ is the sensitivity function of the $V$ magnitude system, and $f_\lambda$ is the monochromatic flux from a star, and $C_2=C_{\rm Bol} - C_{\rm V}$, in which $C_{\rm V}$ is the zero point constant for the $V$ system of magnitudes. This claim, however, was found inapplicable and speculative by \cite{Bakis2022}, who suggested an alternative way of increasing the accuracy of standard $L$ by using multiband standard $BC$s.

\cite{Bakis2022} first determined the main-sequence standard $BC$ values for the photometric bands Gaia $G$, $G_{\rm BP}$, $G_{\rm RP}$ and Johnson $B$, $V$ from the most accurate stellar parameters of 406 main-sequence stars which are the components of 209 DDEBs contained in the catalogue of \cite{Eker2018}. Then, $BC$ - $T_{\rm eff}$ relations of main-sequence stars were established for the same five photometric ($G$, $G_{\rm BP}$, $G_{\rm RP}$ and Johnson $B$, $V$) passbands. After, five $M_{\rm Bol}$ values of each star are estimated by $M_{\rm Bol}(\xi)$ = $M_\xi$ + $BC_\xi$, the mean $M_{\rm Bol}$ and its standard error is computed to represent the star in the list. At last, the standard $L$ of each star is obtained according to Eq.\ref{eq:mbol1} from its mean $M_{\rm Bol}$, while the standard error of $M_{\rm Bol}$ is propagated to be the uncertainty of the predicted $L$. Comparing predicted $L$ (from photometry) to the calculated $L$ (from SB) indicated a high degree of correlation ($R^2$ > 0.999). The most important is that comparing histogram distributions of errors showed that uncertainties associated with the predicted $L$ (peak at $\approx$ 2.5 per cent) are $\sim$3 times smaller than the uncertainties of $L$ (peak at $\approx$ 8 per cent) by the Stefan-Boltzmann law. There was no method providing $L$ more accurately than the direct method. Now, multiband $BC$s provides such an accuracy first in the history of astrophysics.

This study, however, is motivated for further testing of the multiband (Gaia $G$, $G_{\rm BP}$, $G_{\rm RP}$ and Johnson $B$, $V$) main-sequence $BC$ and $BC$ - $T_{\rm eff}$ relations by adding one more passband (TESS) in the series and applying them to the planet-hosting stars with most accurate $T_{\rm eff}$ and $R$. Not only by comparing predicted $L$ to calculated $L$ of the host stars, but also by comparing predicted $R$ to the published $R$. Choosing host stars sample not only among the main-sequence but also subgiants and giants, this study is also important for testing the claim of \cite{Flower1996} who declared $BC$ values are the same for all luminosity classes including main-sequence, subgiants and giants even for the pre-main-sequence stars.

\section{Data}

The Transiting Exoplanet Survey Satellite (TESS) \citep{TESS2014,TESS2015}, which was put into orbit in 2018, has led to important contributions to stellar and planetary astrophysics with its extremely sensitive photometric data \citep[among hundreds of studies;][]{Stassun2017,Fulton2017,Montalto2022}. In addition to its numerous exoplanet discovery, which is the main mission of the satellite, it has also made important contributions to stellar astrophysics \citep[among hundreds of studies;][for different kind of stellar objects]{Gunter2020,Antoci2019,Bakis2022NewA,Espinoza2016}. In the case of eclipsing binaries, it has become possible to obtain relative radii and light contribution of components with a precision better than 0.1 per cent. 

On the way of estimating accurate absolute bolometric magnitudes and luminosity of the component stars of binaries using TESS light curves, interstellar dimming ($A_{\rm TESS}$) and the bolometric corrections ($BC_{\rm TESS}$) are two very critical parameters in addition to a sensitive Gaia distance \citep{Gaia2016}. Despite there is a reliable source of obtaining these two parameters now exist at  Johnson $B$, $V$ and Gaia $G$, $G_{\rm BP}$, $G_{\rm RP}$ band passes (\cite{Bakis2022}), there is no reliable common source satisfying numerous researchers who are interested in solving TESS light curves of eclipsing binaries and exoplanet transiting light curves of  single stars. This study, therefore, is devoted first  to producing an empirical $E(B-V)$ - $A_{\rm TESS}$ relation for estimating $A_{\rm TESS}$ from the $E(B-V)$ colour excess of the system and then to calibrate an empirical main-sequence $BC$ - $T_{\rm eff}$ relation for the TESS pass band for estimating $BC_{\rm TESS}$ from the $T_{\rm eff}$ of the system. Note that, $A_{\rm TESS}$ is also needed in calculating $M_{\rm TESS}$ before obtaining mean $M_{\rm Bol}$ and its uncertainty for a star from six independent estimations of $M_{\rm Bol}(\xi)$ = $M_\xi$ + $BC_\xi$ including the TESS band. The standard $L$ of a star if computed from such a mean $M_{\rm Bol}$ using Eq.\ref{eq:mbol1} has already been shown to produce more accurate stellar $L$ than the classical direct method \citep{Bakis2022} by a sample of DDEB components. Comparing L of the host star sample of this study will be an independent second test of multiband $BC$ - $T_{\rm eff}$ relations. At last, estimating the radii of host stars directly from this luminosity would be a second application tested in this study.

\subsection{Data for establishing \texorpdfstring{$BC_{\rm TESS}$}{BCTESS} - \texorpdfstring{$T_{\rm eff}$}{Teff} relation}

\begin{figure}
\includegraphics[width=\columnwidth]{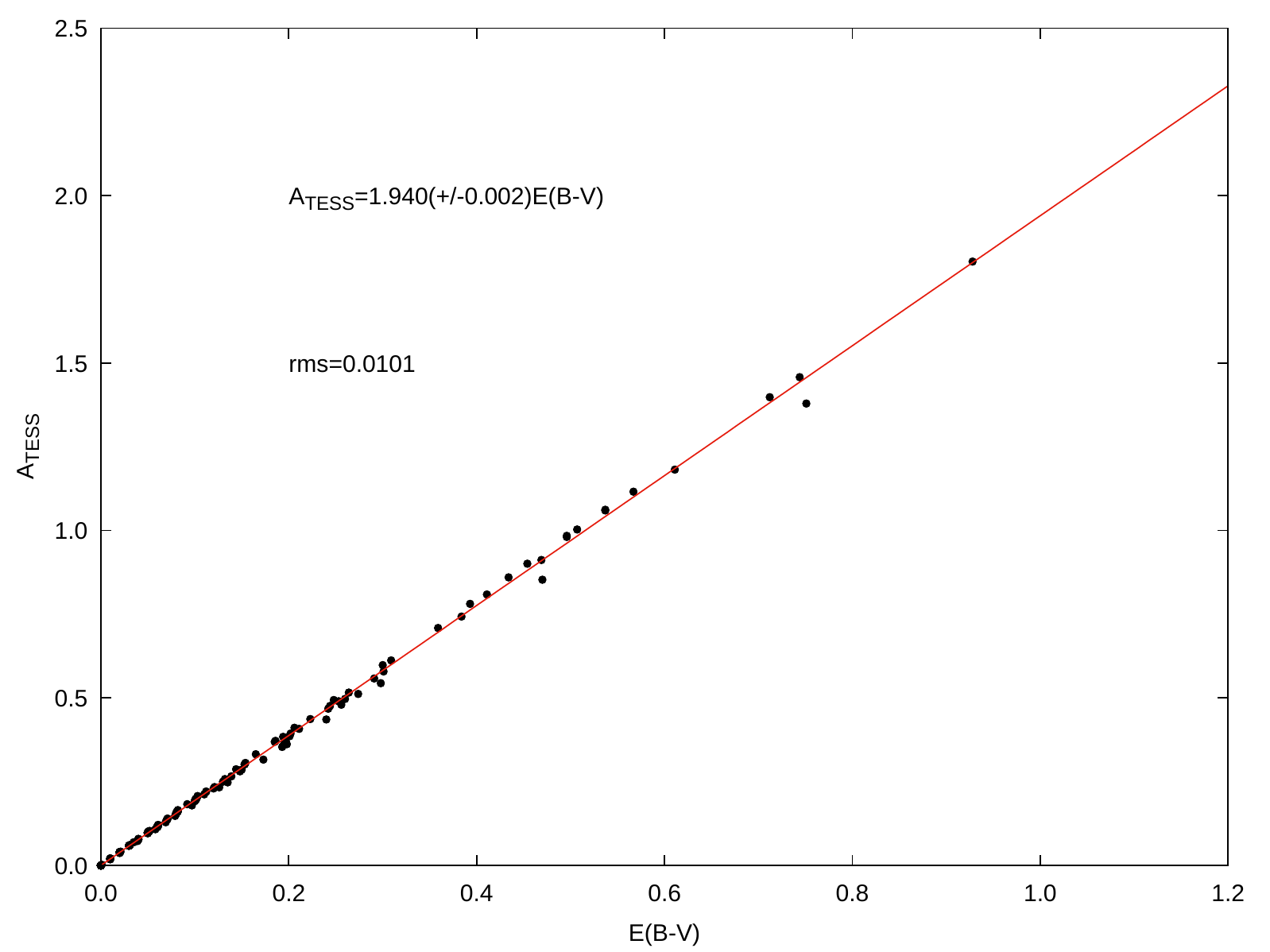}
    \caption{The $E(B-V)$ - $A_{\rm TESS}$ relation derived from 209 DDEB.}  \label{fig:EBV_TESS}
\end{figure}

The same data set of 209 DDEBs from \cite{Bakis2022} are used in this study also for estimating component light ratios ($l$) and interstellar extinction ($A_{\rm TESS}$) first in the TESS magnitudes by the same method and software involving SED and  SIMBAD data as described by \cite{Bakis2022}. After eliminating the stars out of the main-sequence limits, which are the theoretical ZAMS and TAMS lines from PARSEC evolutionary models by \cite{Bressan2012}, the number of component stars left on the main-sequence is 406 (197 binaries, 9 primaries and 3 secondaries). Furthermore, not all of the 209 systems have TESS apparent magnitudes. Therefore, we have been able to use only 390 main-sequence stars to establish $E(B-V)$ - $A_{\rm TESS}$ relation shown in Fig.\ref{fig:EBV_TESS}. It can be used in estimating interstellar dimming in TESS magnitudes if $E(B-V)$ colour excess is known.

\begin{table*}
\centering
\caption{Input parameters and $BC_{\rm TESS}$ values of DDEB stars with TESS apparent magnitudes. The full table is available online.}
\resizebox{\textwidth}{!}
{\begin{tabular}{ccccccccccccccccccccccc}
\hline
Order	&	Name	&	pri/sec	&	$R$ 	& err &	$T$	& err &	$L/L_\odot$ & $L$(SI) & $\frac{\Delta L}{L}$ & $M_{\rm Bol}$ & err & $m({\rm sys})_{\rm TESS}$ & err & $l_{\rm TESS}$ & $m_{\rm TESS}$ & Parallax & $\frac{\sigma_\varpi}{\varpi}$ & $A_{\rm TESS}*$ & $M_{\rm TESS}$ & err & $BC_{\rm TESS}$ & err \\
 & & & [$R_\odot$] & [$\%$] & [$K$] & [$\%$] & & $\times 10^{27}$ & [\%] & [mag] & [mag] & [mag] & [mag] & & [mag] & [mas] & [\%] & [mag] & [mag] & [mag] & [mag] & [mag] \\
\hline
1	&	V421 Peg	&	p	&	1.584	&	1.77	&	7250	&	1.10	&	6.245	&	2.390	&	5.66	&	2.751	&	0.061	&	7.936	&	0.006	&	0.614	&	8.465	&	6.5051	&	0.39	&	0.059	&	2.472	&	0.041	&	0.279	&	0.07	\\
2	&	V421 Peg	&	s	&	1.328	&	2.18	&	6980	&	1.72	&	3.771	&	1.444	&	8.15	&	3.299	&	0.088	&	-	&	-	&	0.386	&	8.970	&	6.5051	&	0.39	&	0.059	&	2.977	&	0.041	&	0.322	&	0.10	\\
3	&	DV Psc	&	p	&	0.685	&	4.38	&	4450	&	0.18	&	0.166	&	0.063	&	8.79	&	6.691	&	0.095	&	9.410	&	0.009	&	0.826	&	9.617	&	23.7216	&	0.09	&	0.370	&	6.123	&	0.041	&	0.569	&	0.10	\\
4	&	DV Psc	&	s	&	0.514	&	3.89	&	3614	&	0.22	&	0.041	&	0.016	&	7.83	&	8.219	&	0.085	&	-	&	-	&	0.174	&	11.311	&	23.7216	&	0.09	&	0.370	&	7.817	&	0.041	&	0.402	&	0.09	\\
5	&	MU Cas	&	p	&	4.192	&	1.19	&	14750	&	3.39	&	749.386	&	286.827	&	13.77	&	-2.447	&	0.149	&	10.549	&	0.020	&	0.555	&	11.187	&	0.5133	&	3.72	&	0.809	&	-1.070	&	0.092	&	-1.377	&	0.18	\\
...	&	...	&	...	&	...	&	...	&	...	&	...	&	...	&	...	&	...	&	...	&	...	&	...	&	...	&	...	&	...	&	...	&	...	&	...	&	...	&	...	&	...	&	...	\\
414	&	AP And	&	s	&	1.195	&	0.44	&	6495	&	2.31	&	2.291	&	0.877	&	9.28	&	3.840	&	0.101	&	-	&	-	&	0.476	&	11.313	&	2.9143	&	0.70	&	0.196	&	3.440	&	0.043	&	0.400	&	0.11	\\
415	&	AL Scl	&	p	&	3.241	&	1.54	&	13550	&	2.58	&	319.014	&	122.103	&	10.78	&	-1.519	&	0.117	&	6.182	&	0.006	&	0.905	&	6.291	&	4.6006	&	3.63	&	0.021	&	-0.416	&	0.089	&	-1.104	&	0.15	\\
416	&	AL Scl	&	s	&	1.401	&	1.43	&	10300	&	3.50	&	19.903	&	7.618	&	14.27	&	1.493	&	0.155	&	-	&	-	&	0.095	&	8.732	&	4.6006	&	3.63	&	0.021	&	2.026	&	0.089	&	-0.533	&	0.18	\\
417	&	V821 Cas	&	p	&	2.308	&	1.21	&	9400	&	4.26	&	37.469	&	14.341	&	17.19	&	0.806	&	0.187	&	8.156	&	0.006	&	0.775	&	8.433	&	3.4262	&	0.59	&	0.061	&	1.047	&	0.042	&	-0.241	&	0.19	\\
418	&	V821 Cas	&	s	&	1.390	&	1.58	&	8600	&	4.65	&	9.522	&	3.644	&	18.87	&	2.293	&	0.205	&	-	&	-	&	0.225	&	9.774	&	3.4262	&	0.59	&	0.061	&	2.388	&	0.042	&	-0.094	&	0.21	\\
\hline
\multicolumn{10}{l}{* Estimated uncertainty is 0.019 mag.}
\end{tabular}}
\label{tableddeb}
\end{table*}

Once, the component light radios and the interstellar extinctions of 209 DDEB systems are known, and then it is straightforward to calculate $BC_{\rm TESS}$ of 390 main-sequence stars as shown by Table \ref{tableddeb}, where the columns are self-explanatory. Order, name of the system and the component (primary or secondary) are given in the first three columns. Only the parameters of 390 stars, a sample selected for this study with TESS magnitudes, are listed thus the numbers in the first column (order) are the same as given by \cite{Bakis2022} for the readers who are interested in looking for the references of the published observed parameters, which are not given in this study to save space; that is, reducing the number of columns in the table. Columns 4, 5, 6 and 7 give radius ($R$), the relative error of $R$, effective temperature ($T_{\rm eff}$) and relative error of $T_{\rm eff}$, from which the luminosity ($L$) of each component is computed according to the Stefan Boltzmann law (column 8) in solar units. It is translated to SI units in column 9. The relative uncertainty of the computed $L$ in column 10 is the propagated uncertainty estimated from the random observational uncertainties of observed $R$ and $T_{\rm eff}$. Absolute bolometric magnitudes ($M_{\rm Bol}$) and corresponding uncertainties in columns 11 and 12 are calculated according to Eq.\ref{eq:mbol1} from the computed $L$. This completes the first step of obtaining $BC_{\rm TESS}$ of the component stars of the DDEB sample (Table \ref{tableddeb}) according to the classical definition $BC_{\rm TESS}$ = $M_{\rm Bol}$ - $M_{\rm TESS}$.

The second step, obtaining $M_{\rm TESS}$ requires apparent brightness of the system $[m({\rm sys})_{\rm TESS}]$, a light ratio of components ($l_{\rm TESS}$), a parallax ($\varpi$) and an interstellar extinction ($A_{\rm TESS}$). Column 13 and 14 retain brightness of the system $[m({\rm sys})_{\rm TESS}]$ and its associated uncertainty $[\Delta m({\rm sys})_{\rm TESS}]$ from the TESS catalogue \citep{Stassun2018}. The light ratio of components (column 15) is calculated as described by \cite{Bakis2022} using the TESS bandpass curve \citep{Sullivan2015} in the form of component contributions ($l_{pri} + l_{sec}= 1$) that is the total contribution of primary and secondary is equal to one. Observational uncertainties $(\Delta m_{\rm TESS})$ of component's magnitudes are assumed to be the same as systemic magnitudes $[\Delta m({\rm sys})_{\rm TESS}]$ (column 14) as done by \cite{Bakis2022}, therefore, $( m_{\rm TESS})$ in column 16 are given without uncertainty in the table. Columns 17 and 18 copy the parallaxes and associated relative errors directly from \cite{Bakis2022} mostly from EDR3, which are checked and confirmed to be the same as Gaia DR3 parallaxes. Column 19 presents interstellar extinctions for the TESS photometry ($A_{\rm TESS}$) predicted in this study with an estimated accuracy of $0.019$ mag. Next, the absolute magnitude of a component ($M_{\rm TESS}$) and its propagated uncertainty are listed in columns 20 and 21. At last in the third step, the $BC_{\rm TESS}$ of a component (column 22) is found by subtracting $M_{\rm TESS}$ (column 20) from $M_{\rm Bol}$ (column 11). The last column (column 23) gives the propagated uncertainty computed from the uncertainties of the absolute bolometric and TESS magnitudes (column 23).

\begin{figure}
\includegraphics[width=\columnwidth]{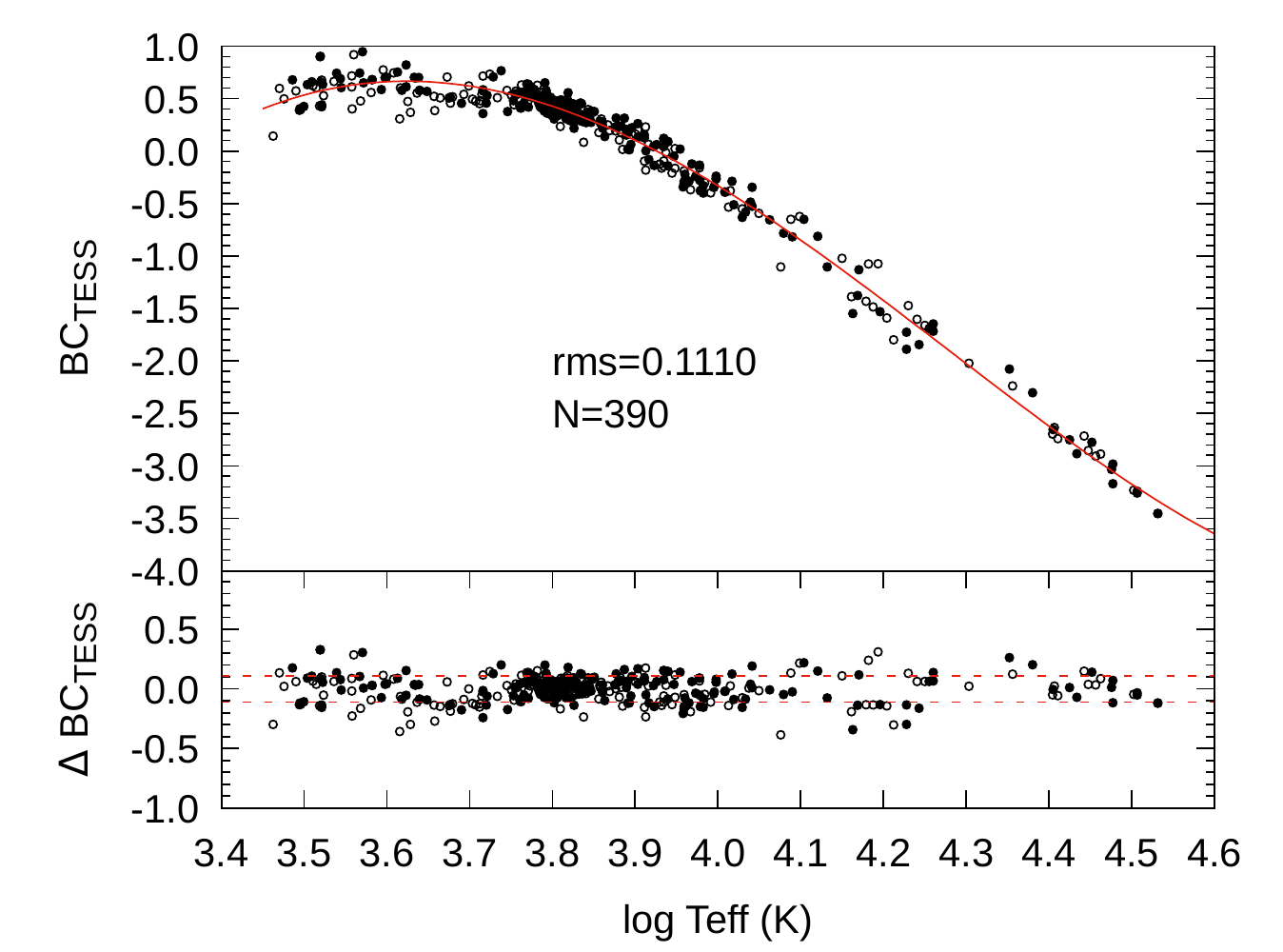}
    \caption{Distribution of empirical standard $BC_{\rm TESS}$ from 209 DDEB. Filled and empty circles are primaries and secondaries, respectively. The best fitting line represents empirical standard $BC_{\rm TESS}$ - $T_{\rm eff}$ relation. The number of stars ($N$) and RMS are indicated. One sigma deviation from the best-fitting curve is shown below.} \label{fig:BC_TESS}
\end{figure}

\subsection{Data for expanding \texorpdfstring{$BC$}{BC} - \texorpdfstring{$T_{\rm eff}$}{Teff} relations}

Empirical standard $BC_{\rm TESS}$  from the published $R$ and $T_{\rm eff}$ of 390 DDEB main-sequence components are displayed in Fig.\ref{fig:BC_TESS} where the continuous line is the best fitting curve. Coefficients and associated uncertainties are listed in Table \ref{tab:BCpartable} together with other $BC - T_{\rm eff}$ relations all in the forms of fourth-degree polynomials from \cite{Bakis2022} representing the photometric bands of Johnson $B$, $V$, Gaia $G$, $G_{\rm BP}$ and $G_{\rm RP}$, where the uncertainties of the coefficients are shown by $\pm$ symbol. We have tried to fit both of the third and fourth-degree polynomials to the $BC_{\rm TESS}$ of this study, nevertheless, the fit of the third-degree polynomial is found better with a smaller RMS and more meaningful errors associated with the coefficients while the curves of both functions are very similar. Using the values of $a$, $b$, $c$, $d$ and $e$ from the table, one can calculate 

\begin{equation}
    BC_{\xi}=a+bX+cX^2+dX^3+eX^4,
	\label{eq:BC_Teff}
\end{equation}
of any photometric band, where $X=log\, T_{\rm eff}$. The lower part of Table \ref{tab:BCpartable} is for comparing standard deviations (RMS), correlations ($R^2$) and the standard $BC$ of a main sequence star having a $T_{\rm eff}$ similar to the Sun ($T_{\rm eff}$ = 5772 K). The smallest RMS, which is 0.1092 mag, occurs at Gaia $B_{\rm RP}$ band, while the largest RMS, which is 0.1363 mag, occurs at Johnson $B$ band. The RMS of the TESS band has a moderate value (0.1110 mag) in between these limits. Note that the RMS values of each band determine the limiting accuracy of $M_{\rm Bol}$ sınce $M_{\rm Bol}$ = $M_\xi + BC_\xi$ under the condition the uncertainty of $M_\xi$ is negligible. The Maximum $BC$ values ($BC_{\rm max}$) occurring at effective temperature $T_{\rm max}$ are given below the absolute and apparent magnitudes if the main-sequence star with $T_{\rm eff}$ = 5772 K is  shown by symbols $M_\odot$ and $m_\odot$ in the table. The lowest part of the table indicates the range of positive $BC$ values if exist. Accordingly, the $BC$ values of Johnson $B$ and Gaia $G_{\rm BP}$ bands are always negative. The $BC$ values of Johnson $V$ and Gaia $G$ bands would produce positive $BC$ values in the middle temperatures; for the $V$ band if 5300 < $T_{\rm eff}$ < 7830 K, for the G band if 4565 < $T_{\rm eff}$ < 4720 K.  The $BC$ values of Gaia $G_{\rm RP}$ are positive if $T_{\rm eff}$ < 5890 K, while the $BC_{\rm TESS}$ are positive if $T_{\rm eff}$ < 8425 K. The relations are set valid for the main-sequence stars within the temperature range 2900-38000 K as it is implied by the sample of DDEB used in the calibrations.

\begin{figure}
\includegraphics[width=\columnwidth]{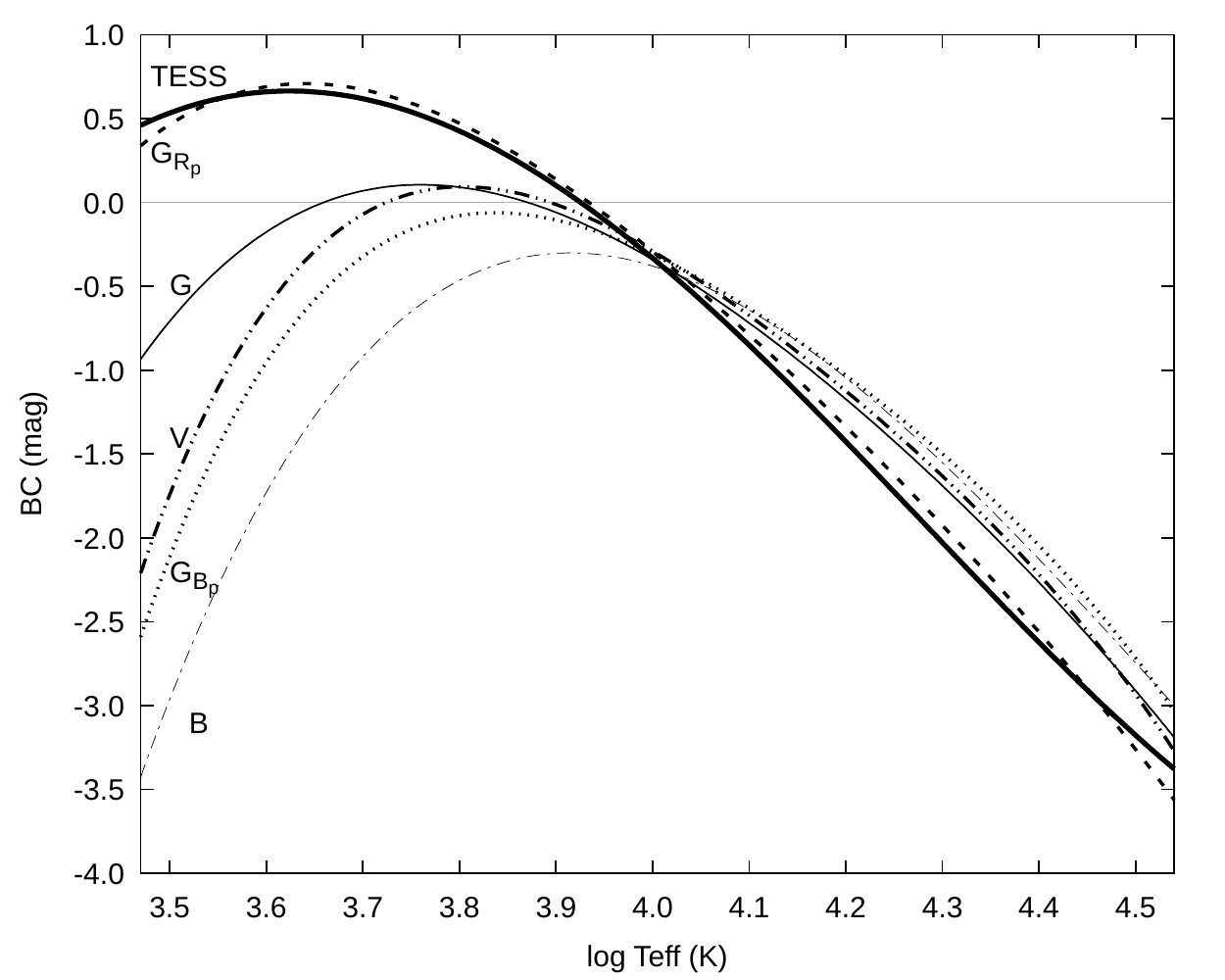}
    \caption{Independently calibrated six-band standard $BC$ - $T_{\rm eff}$ relations. The names of the passbands are indicated.} \label{fig:allBCs}
\end{figure}

Independently calibrated multiband standard $BC$ - $T_{\rm eff}$ relations are plotted all together in Fig.\ref{fig:allBCs}. Except $TESS$ (thick solid) and $G_{\rm RP}$ (dashed) curves, which are similar, all of the other curves deviate from each other, especially towards lowest temperatures while all curves cross each other at $T_{\rm eff}$$\approx$10000 K and appear not deviating from each other $T_{\rm eff}>$10000 K as much as lower temperatures.  

Each curve in Fig.\ref{fig:allBCs} could be used to estimate a $BC$ of a star at a photometric band preferred if its effective temperature were known. Because $BC$ - $T_{\rm eff}$ relations are independently calibrated and because observations at different photometric bands are independent, the predicted values of $M_{\rm Bol}$ of various bands would be independent. The six independent $M_{\rm Bol}$ for a single star could then be combined by taking an average. At last, the standard $L$ of a star is calculated from the averaged $M_{\rm Bol}$. Otherwise, each $M_{\rm Bol}$ providing a standard $L$ would be less accurate than the limiting accuracy indicated by the RMS values in Table \ref{tab:BCpartable}. 

\begin{table}
\centering
\caption{Coefficients of the fourth-degree polynomials ($BC$ - $T_{\rm eff}$ relations) for estimating $BC_{\xi}$, where $\xi$ could be one of the Johnson $B$, $V$, Gaia $G$, $G_{\rm BP}$, $G_{\rm RP}$ and TESS bands }
\resizebox{\columnwidth}{!}{\begin{tabular}{cccccccc}
\hline
Coefficient & $BC_{\rm B}^{*}$ & $BC_{\rm V}^{*}$ & $BC_{G}^{*}$ & $BC_{G_{\rm BP}}^{*}$ & $BC_{G_{\rm RP}}^{*}$ & $BC_{\rm TESS}$\\
\hline
a & --1272.43 & --3767.98 & --1407.14 & --3421.55 & --1415.67 &  --318.533 \\
  & $\pm$394.2 & $\pm$288.8 & $\pm$256.7 & $\pm$293.6 & $\pm$253.3 & $\pm$18.92 \\
b & 1075.85   & 3595.86  & 1305.08  & 3248.19 & 1342.38 &  232.298 \\
  & $\pm$394.4 & $\pm$290.9 & $\pm$258.9 & $\pm$296.1 & $\pm$255.4 & $\pm$14.4 \\
c & --337.831 & --1286.59 & --453.605 & --1156.82 & --475.827 &  --55.2916\\
  & $\pm$147.7 & $\pm$109.6 & $\pm$97.67 & $\pm$111.7 & $\pm$96.34 & $\pm$3.642\\
d & 46.8074 & 204.764  & 70.2338 & 183.372 & 74.9702 & 4.27613\\
  & $\pm$24.53 & $\pm$18.32 & $\pm$16.34 & $\pm$18.68  & $\pm$16.12  & $\pm$0.3061\\
e & --2.42862 & --12.2469 & --4.1047 & --10.9305 & --4.44923 & 0\\
  & $\pm$1.552 & $\pm$1.146  & $\pm$1.023 & $\pm$1.169   & $\pm$1.009 & -\\
\hline
rms & 0.136257 & 0.120071 & 0.11068 & 0.126577 & 0.109179 & 0.110951 \\
$R^2$ & 0.9616 & 0.9789 & 0.9793 & 0.9738 & 0.9884 & 0.9847\\
$BC_\odot$ (mag) & --0.600 & 0.069 & 0.106 & --0.134 & 0.567 & 0.517\\
$M_\odot$ (mag) & 5.340 & 4.671 & 4.634 & 4.874 & 4.173 & 4.223 \\
$m_\odot$ (mag) & --26.232 & --26.901 & --26.938 & --26.698 & --27.399 & --27.349\\
\hline
$BC_{\rm max}$ (mag) & --0.301 & 0.094 & 0.106 & --0.062 & 0.709 & 0.664 \\
$T_{\rm max}$ (K) & 8222 & 6397 & 5715 & 6879 & 4345 & 4210 \\
\hline
$T_{1}$ (K) &-& 5300 & 4565 &-&-& -\\
$T_{2}$ (K) &-& 7830 & 7420 &-&8590 & 8425\\
\hline
*: \cite{Bakis2022}.
\end{tabular}}
\label{tab:BCpartable}
\end{table}

\subsection{Data for Testing \texorpdfstring{$BC$}{BC} - \texorpdfstring{$T_{\rm eff}$}{Teff} Relations}

Five band ($B$, $V$, $G$, $G_{\rm RP}$ and $G_{\rm BP}$) $BC$ and $BC$ - $T_{\rm eff}$ relations has already been tested by \cite{Bakis2022} by a sample of DDEB components from which $BC_\xi$ were computed and $BC_\xi$ - $T_{\rm eff}$ calibrated, where $\xi$ is any of the five photometric bands. Simply because DDEB are known to provide the most accurate stellar parameters \citep{Andersen1991,Torres2010a,Eker2014, Eker2015,Eker2018} and are already ready for such a test. The next most accurate stellar parameters appear to be coming from single stars which are hosting one or more exoplanets discovered by radial velocity variations and/or transiting light curves. High resolution and high signal-to-noise ratio spectra provide reliable $log\,g$ and $T_{\rm eff}$ of the host stars, while relatively shallow transits detected by ultra-high signal to ration light curves like TESS, on the other hand, resemble eclipses of DDEB, from which stellar $R$ and $M$ could be estimated at about $\sim8$ per cent and 30 per cent \citep{Stassun2017} using the direct observables. Later, the method is revised to provide host star $R$ and $M$ at a level of 10 per cent and 25 per cent by \cite{Stassun2018}. At last, with the improved parallaxes of DR2 at the time and using granulation based $log\,g$ via Fourier background modelling \citep{Corsaro2017} with TESS, the accuracy reached to 3 and 10 per cent levels respectively, which appear ideal for testing existing multiband  $BC$ and $BC$ - $T_{\rm eff}$ relations by a different sample of stars other than DDEB components.

Host star physical parameters and associated errors are listed in NASA Exoplanet Archive\footnote{\url{https://exoplanetarchive.ipac.caltech.edu/}} together with the physical and orbital parameters of the confirmed planets.  There was 5081 confirmed exoplanet belonging to 3799 host stars at the time when we downloaded the planetary systems composite data, from which we have selected 306 host stars having the most accurate $R$ and $T_{\rm eff}$ both within 2 per cent of accuracy for our preliminary list. This way, host stars with the most accurate stellar $L$ at all luminosity classes were collected intentionally in order to see how good existing main-sequence multiband $BC$ - $T_{\rm eff}$ relations are usable to other stages of evolution. Noting that the $L$ of a star primarily depends on its mass and metallicity, which is critical to identify sub-giants and main-sequence stars on H-R diagram, the host stars with a published metallicity were preferred. Because of the note “Data may not be self-consistent if drawn from multiple sources” on the table of Planetary Systems Composite Data, we have studied each star from its original sources to make sure self-consistency. In the second step, where we had to replace some of the original choices of $R$ and $T_{\rm eff}$ with a new $R$ and $T_{\rm eff}$ together with their associated uncertainty for the sake of consistency. We preferred not to discard a host star if the uncertainty of the new entry is bigger than 2 per cent but less than 3 per cent. Additional host stars fitting the selection criteria were also added thus our final list is enlarged to have 350 host stars.     

However, during the process of estimating interstellar dimming of the selected host stars one by one from their un-reddened and reddened SED fitting, the following six systems  KELT-21, WASP-118, HATS-37, Kepler-38, Kepler-1647 and Kepler-1661 were discarded because they have secondaries polluting their SED. The bright host $\alpha$ Tau was also discarded because it does not have Gaia apparent magnitudes. We had no choice but to discard the host K2-374 because its observed spectrophotometric fluxes could not be reached and the host K2-127 was discarded because its observed spectrophotometric fluxes were found not to fit its SED.

The sample containing 341 hosts selected for this study for testing multiband $BC$ and $BC$ - $T_{\rm eff}$ relations are listed in Table \ref{tablehostphysics}. Order and the star name are in columns 1 and 2 while parallax, error and reference are in columns 3, 4, and 5. Columns 6, 7 and 8 indicate radius, radius error and reference while columns 9, 10, 11 and 12 show spectral type, $T_{\rm eff}$, temperature error and reference. Column 13 is the luminosity class determined by us using ZAMS and TAMS lines of PARSEC models \citep{Bressan2012} according to the published metallicity. The luminosity according to the Stefan-Boltzmann law from the observed $R$ and $T_{\rm eff}$ in solar and SI units are in columns 14 and 15 respectively while its propagated relative uncertainty ($\Delta L/L$) is in column 16. Corresponding $M_{\rm Bol}$ and uncertainty in the magnitude scale according to Eq.\ref{eq:mbol1} are given in columns 17 and 18.

\begin{figure}
\includegraphics[width=\columnwidth]{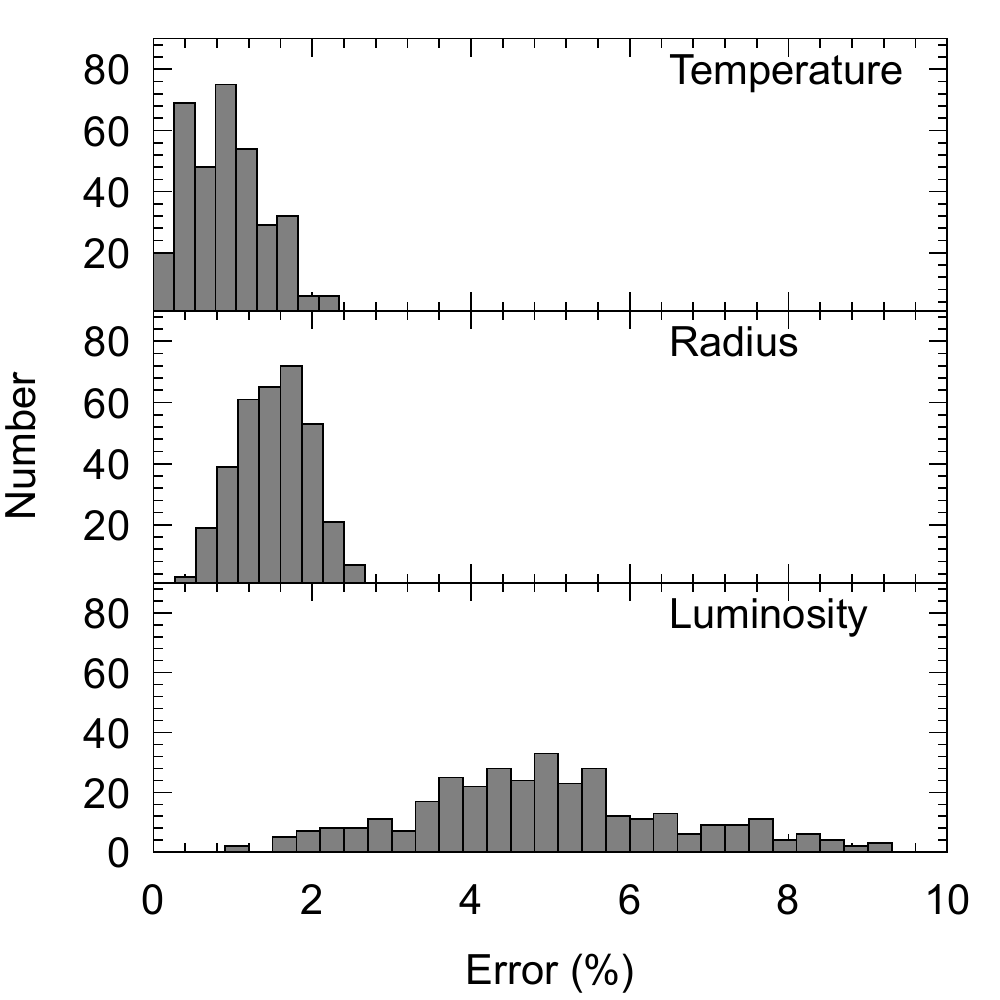}
    \caption{Error distributions of effective temperatures, radii and luminosity of  341 single host-stars in the present sample.} \label{fig:histLRT}
\end{figure}

\begin{figure}
\includegraphics[width=\columnwidth]{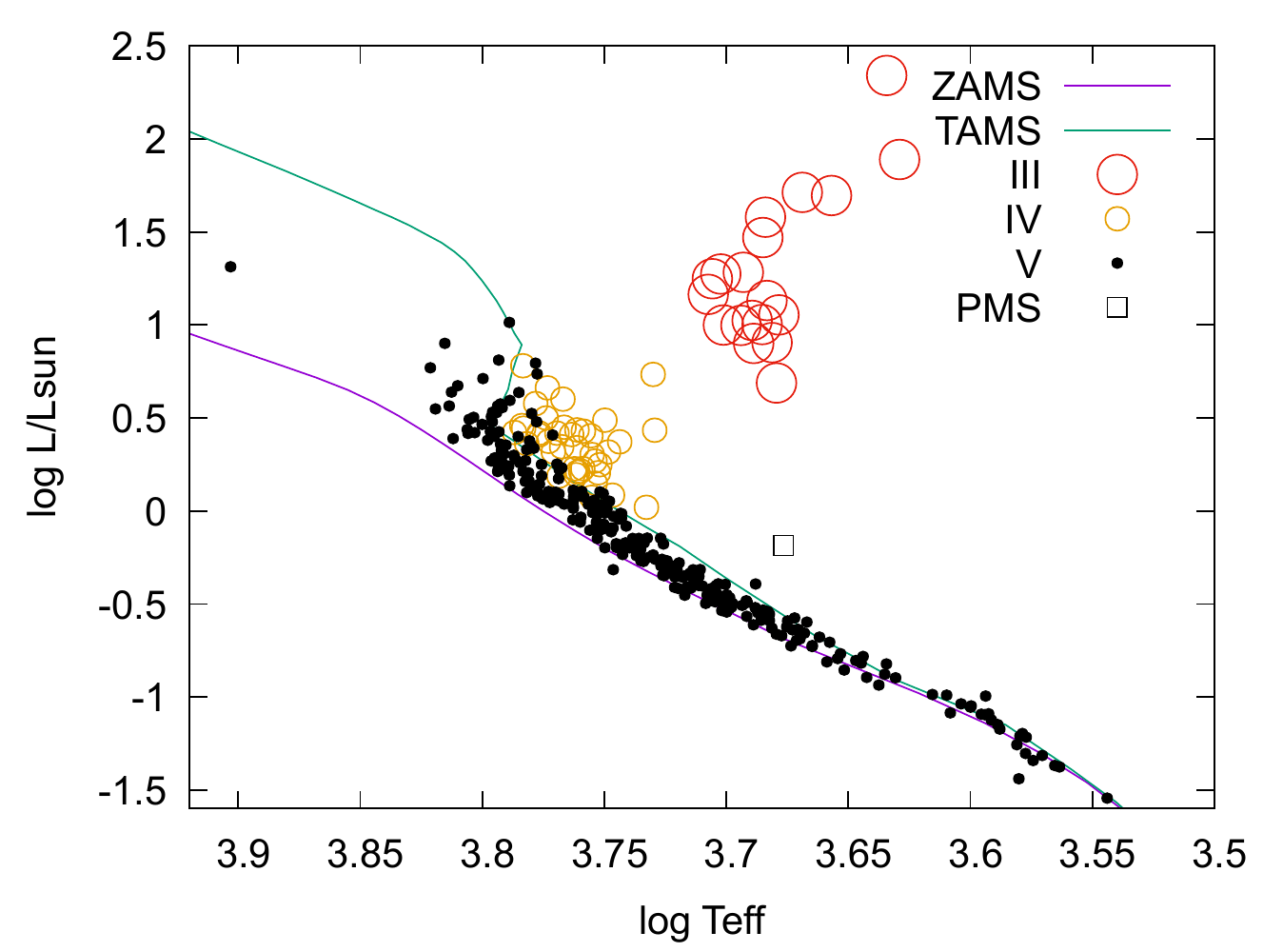}
    \caption{HR diagram of the host stars sample. The main sequence limits indicated by ZAMS and TAMS of the solar metallicity (Z=0.014) from PARSEC models \citep{Bressan2012} while large empty circles are giants, medium empty circles are sub-giants, small filled circles are main-sequence stars and the empty square is a pre-main-sequence star. }  \label{fig:HR}
\end{figure}

Observational random errors of $R$ and $T_{\rm eff}$ are compared to the propagated errors of $L$ in Fig.\ref{fig:histLRT}, where a much wider distribution of $L$ errors with a peak at 5 per cent caused by the powers of $R$ and $T_{\rm eff}$ is obvious.  Fig.\ref{fig:HR} displays the positions of the sample host stars on the H-R diagram where the luminosity classes are indicated. Thus, the single host star sample contains 281 main-sequence (V), 40 sub-giants (IV), 19 giants and one pre-main-sequence (PMS) star YSES-2 \citep{Bohn2021}.

\begin{table*}
\centering
\caption{The sample of host stars with most accurate radii and effective temperatures chosen for testing $BC$ and $BC - T_{\rm eff}$ relations.}
\resizebox{\textwidth}{!}
{\begin{tabular}{cccccccccccccccccc}
\hline
Order	&	Host	&	Parallax	&	$\frac{\sigma_\varpi}{\varpi}$	&	Plx	&	$R$	&	err	&	Reference	&	Sp. & $T$	&	err	&	Reference	&	Lum.	&	$L/L_\odot$	&	$L$(SI)	&	$\frac{\Delta L}{L}$	&	$M_{\rm Bol}$	&	err	\\
&	Name	&	[mas]	&	[\%]	&	Source	&	[$R_\odot$]	&	[$R_\odot$]	&		& Type &	[K]	&	[K]	&		&	Class	&		&	$\times 10^{27}$	&	[\%]	&	[mag]	&	[mag]	\\
\hline
1	&	16 Cyg B	&	47.3302	&	0.0362	&	DR3	&	1.13	&	0.01	&	2017AJ....153..136S	&	G3 V	&	5750	&	8	&	2017AJ....153..136S	&	V	&	1.258	&	0.481	&	1.9	&	4.491	&	0.020	\\
2	&	24 Sex	&	13.6402	&	0.1960	&	DR3	&	4.90	&	0.08	&	2011AJ....141...16J	&	K0 IV	&	5098	&	44	&	2011AJ....141...16J	&	III	&	14.611	&	5.592	&	4.8	&	1.828	&	0.052	\\
3	&	2MASSJ08421149+1916373	&	5.4387	&	0.3377	&	DR3	&	0.83	&	0.01	&	2016A\&A...588A.118M	&	0	&	5300	&	30	&	2016A\&A...588A.118M	&	V	&	0.486	&	0.186	&	3.7	&	5.523	&	0.040	\\
4	&	55 Cnc	&	79.4482	&	0.0540	&	DR3	&	0.96	&	0.02	&	2021ApJS..255....8R	&	K0 IV-V	&	5318	&	81	&	2021ApJS..255....8R	&	V	&	0.664	&	0.254	&	7.4	&	5.185	&	0.080	\\
5	&	61 Vir	&	117.1726	&	0.1243	&	DR3	&	0.96	&	0.01	&	2010ApJ...708.1366V	&	G5 V	&	5577	&	33	&	2010ApJ...708.1366V	&	V	&	0.808	&	0.309	&	3.3	&	4.971	&	0.036	\\
…	&	…	&	…	&	…	&	…	&	…	&	…	&	…	&	…	&	…	&	…	&	…	&	…	&	…	&	…	&	…	&	…	&	…	\\
337	&	WASP-187	&	2.7780	&	0.6339	&	DR3	&	2.83	&	0.05	&	2020MNRAS.499..428S	&	F	&	6150	&	89	&	2020MNRAS.499..428S	&	V	&	10.322	&	3.951	&	6.8	&	2.206	&	0.073	\\
338	&	WASP-189	&	10.0997	&	0.2901	&	DR3	&	2.36	&	0.03	&	2020A\&A...643A..94L	&	A	&	8000	&	80	&	2020A\&A...643A..94L	&	V	&	20.553	&	7.867	&	4.7	&	1.458	&	0.051	\\
339	&	XO-2 N	&	6.6588	&	0.2367	&	DR3	&	0.99	&	0.01	&	2012ApJ...761....7C	&	K0 V	&	5332	&	57	&	2015A\&A...575A.111D	&	V	&	0.714	&	0.273	&	4.6	&	5.106	&	0.050	\\
340	&	XO-7	&	4.3216	&	0.3066	&	DR3	&	1.48	&	0.02	&	2020AJ....159...44C	&	G0 V	&	6250	&	100	&	2020AJ....159...44C	&	V	&	3.011	&	1.153	&	7.1	&	3.543	&	0.077	\\
341	&	YSES 2	&	9.1537	&	0.1291	&	DR3	&	1.19	&	0.02	&	2021A\&A...648A..73B	&	K1 V	&	4749	&	40	&	2021A\&A...648A..73B	&	PMS	&	0.652	&	0.250	&	5.0	&	5.204	&	0.054	\\
\hline
\end{tabular}}
\label{tablehostphysics}
\end{table*}

\section{Calculations}

\subsection{Estimating interstellar extinction using SED}

Assuming a star is a black body having a temperature $T_{\rm eff}$ and if there is no interstellar extinction, a spherical star of size $R$ would produce a continuum flux (SED) at a distance $d$ from its centre. Putting this star at the $d$ parsec away from the Earth, according to the notation of \cite{Bakis2022}, its monochromatic flux above the atmosphere is  
\begin{equation}
    f_\lambda = \frac{R^2}{d^2} \pi B_\lambda(T_{\rm eff}).
	\label{eq:flambda}
\end{equation}
where 
$\pi B_\lambda(T_{\rm eff})$ is monochromatic surface flux and $R^2$/$d^2$ is the dilution factor.

\begin{figure}
  \centering
  \includegraphics[width=1.\columnwidth]{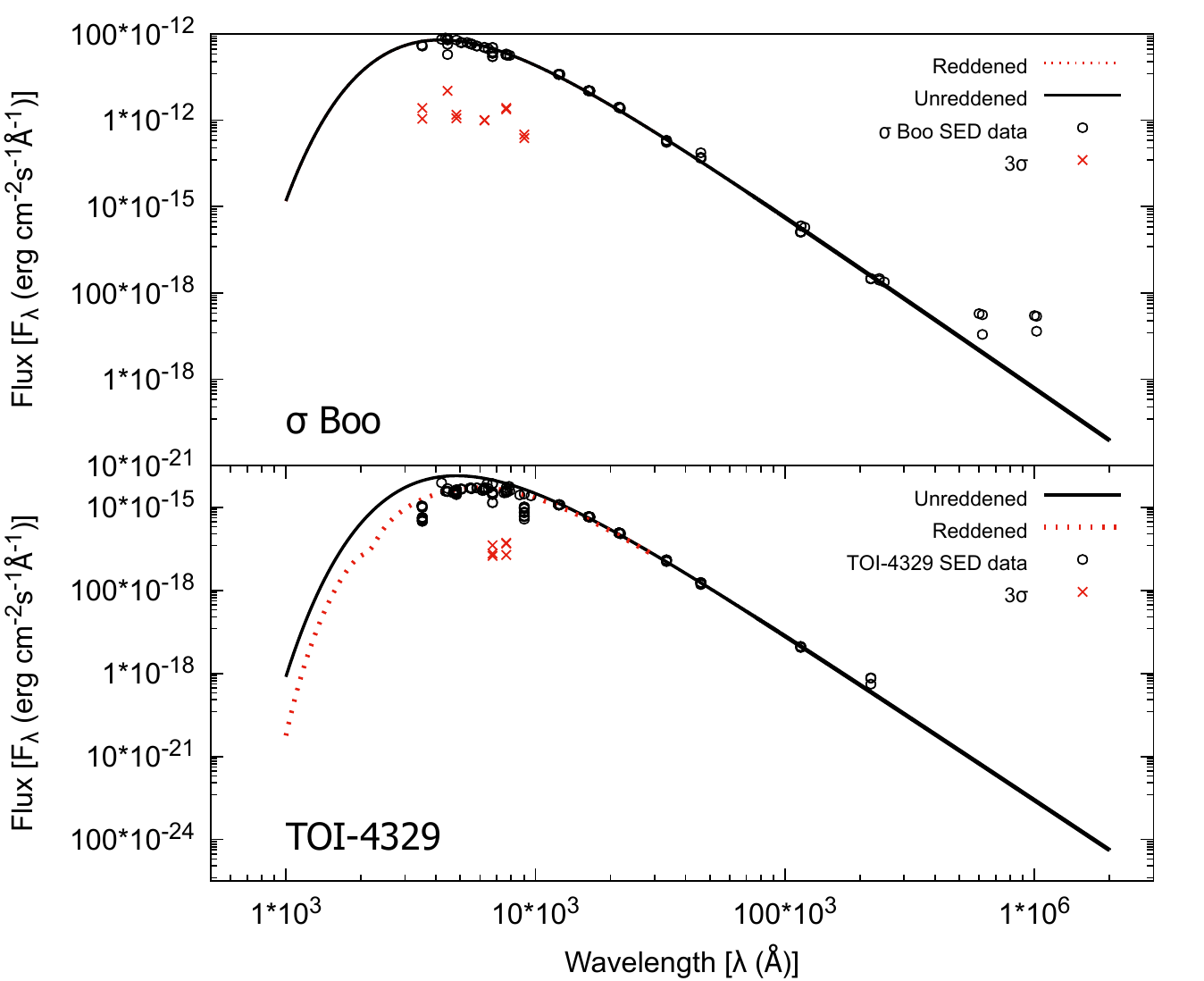}
\caption{SED data and model for $\sigma$ Boo (\textit{top}) and TOI 4329 (\textit{bottom}). The data refers to within 5 arcsecs around the objects.}
\label{fig:SED}
\end{figure}

Using the parallaxes ($\varpi$), radii ($R$) and effective temperatures ($T_{\rm eff}$) in Table \ref{tablehostphysics}, the unreddened SED of the host stars in this study are computed in units of $W m^{-2}$ \AA$^{-1}$ and compared to their observed spectrophotometric flux data from the SIMBAD database \citep{simbad2000}. For the nearby hosts with no interstellar extinction, like $\sigma$ Boo in the \textit{upper} panel of Fig.\ref{fig:SED}, the SED are found to fit perfectly to observed spectrophotometric flux data at all wavelengths. But, for the hosts with interstellar extinction like TOI-4329, the observed data towards the short wavelengths would be found off while the fit towards the long wavelengths appears acceptable to confirm the observed input parameters ($\varpi, R$ and $T_{\rm eff}$ ) are consistent. Inconsistency occurs if one or all of the observed parameters are determined wrongly and/or there is excess radiation in the system, which could be due to radiating circumstellar dust or flux from a companion star if the host is not a single star but a close binary. 

The reddened SED of the host stars is modelled one by one by adjusting $E(B-V)$ of the system until a best-fitting reddened SED is obtained using the reddening model of \cite{Fitzpatrick1999}. Unreddened SED and best fitting reddened SED of the host star TOI-4329 are shown together in the \textit{lower} panel of Fig.\ref{fig:SED}.

\begin{figure}
\includegraphics[width=\columnwidth]{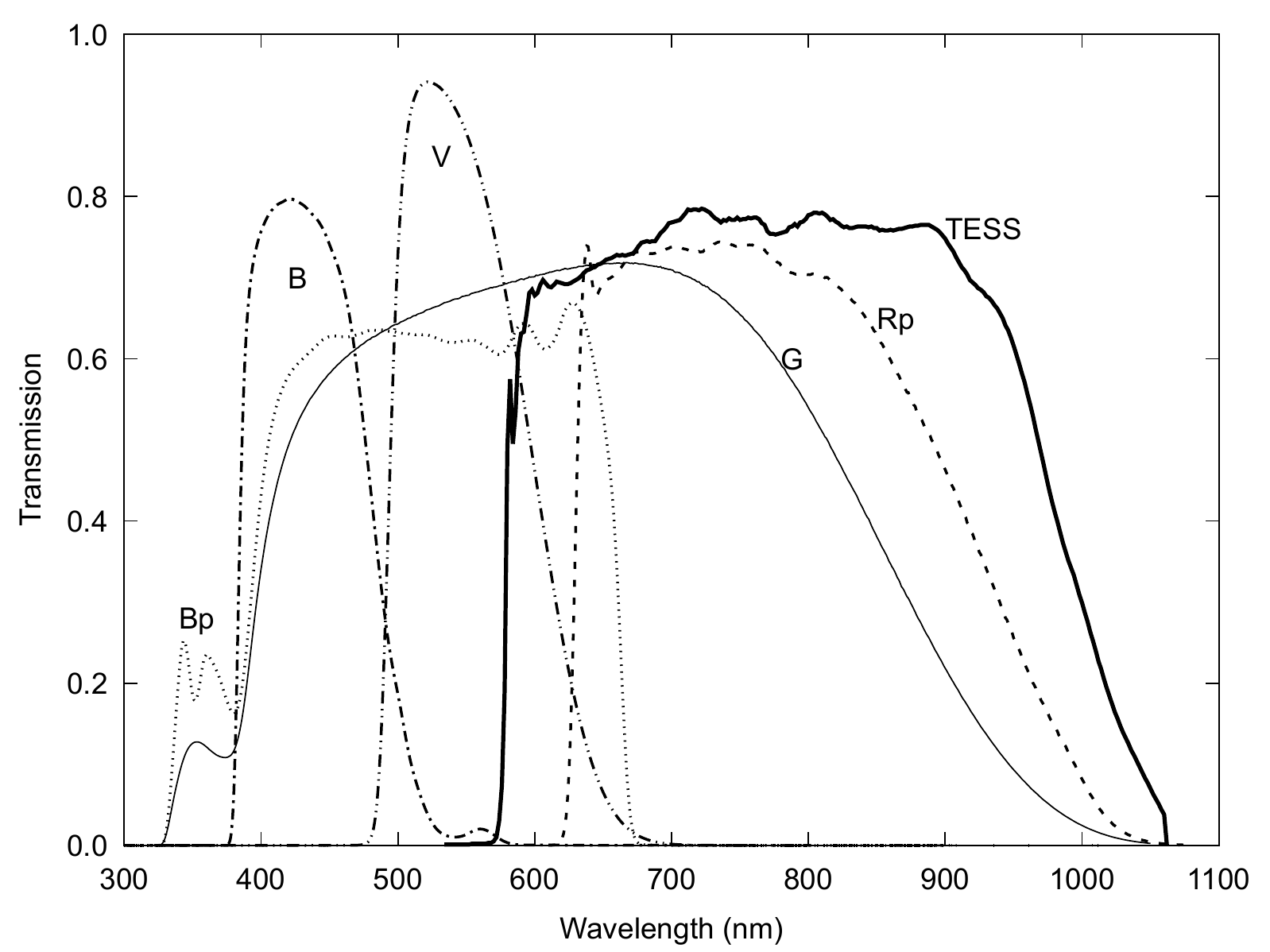}
    \caption{Transmission curves of photometric passband used in SED analysis.}  \label{fig:filters}
\end{figure}

Filter transition profiles, $S_\lambda(\xi)$, of the photometric bands Johnson $B$, $V$, Gaia $G$, $G_{\rm BP}$, $G_{\rm RP}$ and $TESS$ are needed to calculate interstellar extinctions, $A_\xi$,

 \begin{equation}
    A_\xi = 2.5log\frac{\int_{0}^{\infty} S_\lambda(\xi)f_\lambda^0 d\lambda}{\int_{0}^{\infty} S_\lambda(\xi)f_\lambda d\lambda},
	\label{eq:A_dimming}
\end{equation}
where it is clear that if un-reddened  $f_\lambda^0$ and reddened $f_\lambda$ are the same, $E(B-V)$ and all $A_\xi$ would be zero, which means no interstellar extinction. The filter profiles of the photometric bands are displayed in Fig.\ref{fig:filters} where transmission data for Johnson $B$ and $V$ are taken from \cite{Bessel1990}, Gaia passbands from \cite{Evans2018} and TESS from \cite{Sullivan2015}. The passband-based interstellar extinctions determined by equation 6 for the host stars in this study are given in Table \ref{tablehostmags} together with apparent magnitudes. 


\begin{table*}
\centering
\caption{Apparent magnitudes and extinctions of the host stars sample at Johnson $B$, $V$, Gaia $G$, $G_{\rm BP}$, $G_{\rm RP}$ and TESS passbands. The full table is available online.}
\resizebox{\textwidth}{!}
{\begin{tabular}{cccccccccccccccccccc}
\hline
Order	&	Hostname	&	$B$	&	err	&	$V$	&	err	&	$G$	&	err	&	$G_{\rm BP}$	&	err	&	$G_{\rm RP}$	&	err	&	$TESS$	&	err	&	$A_{\rm B}*$	&	$A_{\rm V}*$	&	$A_G*$	&	$A_{GBP}*$	&	$A_{GRP}*$	&	$A_{\rm TESS}*$	\\
	&		&	[mag]	&	[mag]	&	[mag]	&	[mag]	&	[mag]	&	[mag]	&	[mag]	&	[mag]	&	[mag]	&	[mag]	&	[mag]	&	[mag]	&	[mag]	&	[mag]	&	[mag]	&	[mag]	&	[mag]	&	[mag]	\\
 \hline
1	&	16 Cyg B	&	6.876	&	0.072	&	6.215	&	0.016	&	6.073	&	0.003	&	6.401	&	0.003	&	5.575	&	0.004	&	5.628	&	0.006	&	0.000	&	0.000	&	0.000	&	0.000	&	0.000	&	0.000	\\
2	&	24 Sex	&	7.399	&	0.019	&	6.454	&	0.023	&	6.215	&	0.003	&	6.676	&	0.003	&	5.584	&	0.004	&	5.642	&	0.006	&	0.121	&	0.091	&	0.080	&	0.099	&	0.056	&	0.057	\\
3	&	2MASSJ08421149+1916373	&	13.062	&	0.005	&	12.163	&	0.080	&	11.926	&	0.003	&	12.360	&	0.003	&	11.325	&	0.004	&	11.385	&	0.006	&	0.162	&	0.122	&	0.107	&	0.132	&	0.075	&	0.077	\\
4	&	55 Cnc	&	6.816	&	0.019	&	5.951	&	0.023	&	5.733	&	0.003	&	6.156	&	0.003	&	5.147	&	0.004	&	5.206	&	0.006	&	0.000	&	0.000	&	0.000	&	0.000	&	0.000	&	0.000	\\
5	&	61 Vir	&	5.440	&	0.007	&	4.740	&	0.009	&	4.533	&	0.003	&	4.905	&	0.003	&	4.006	&	0.006	&	4.085	&	0.008	&	0.000	&	0.000	&	0.000	&	0.000	&	0.000	&	0.000	\\
…	&	…	&	…	&	…	&	…	&	…	&	…	&	…	&	…	&	…	&	…	&	…	&	…	&	…	&	…	&	…	&	…	&	…	&	…	&	…	\\
337	&	WASP-187	&	10.836	&	0.083	&	10.295	&	0.006	&	10.138	&	0.003	&	10.449	&	0.003	&	9.657	&	0.004	&	9.712	&	0.010	&	0.243	&	0.183	&	0.168	&	0.203	&	0.114	&	0.117	\\
338	&	WASP-189	&	6.788	&	0.025	&	6.598	&	0.023	&	6.572	&	0.003	&	6.658	&	0.003	&	6.388	&	0.004	&	6.416	&	0.006	&	0.000	&	0.000	&	0.000	&	0.000	&	0.000	&	0.000	\\
339	&	XO-2 N	&	11.947	&	0.161	&	11.246	&	0.011	&	10.971	&	0.003	&	11.390	&	0.003	&	10.392	&	0.004	&	10.447	&	0.006	&	0.000	&	0.000	&	0.000	&	0.000	&	0.000	&	0.000	\\
340	&	XO-7	&	11.229	&	0.087	&	10.521	&	0.006	&	10.460	&	0.003	&	10.764	&	0.003	&	9.999	&	0.004	&	10.052	&	0.006	&	0.284	&	0.214	&	0.196	&	0.238	&	0.133	&	0.137	\\
341	&	YSES 2	&	11.767	&	0.151	&	10.885	&	0.011	&	10.525	&	0.003	&	11.030	&	0.008	&	9.862	&	0.007	&	9.932	&	0.006	&	0.201	&	0.152	&	0.130	&	0.163	&	0.093	&	0.095	\\
\hline
\multicolumn{20}{l}{* Uncertainties: 0$^m$.04 ($A_{\rm B}$), 0$^m$.03 ($A_{\rm V}$), 0$^m$.029 ($A_{G}$), 0$^m$.035 ($A_{G_{\rm BP}}$), 0$^m$.019 ($A_{G_{\rm RP}}$) and 0$^m$.019 ($A_{\rm TESS}$).}
\end{tabular}}
\label{tablehostmags}
\end{table*}

\subsection{Calculating standard \texorpdfstring{$L$}{L} by multiband \texorpdfstring{$BC$}{BC}}

Using the apparent magnitudes and the interstellar extinctions from Table \ref{tablehostmags} and the parallaxes from Table \ref{tablehostphysics},  the multiband absolute magnitudes of the host stars are calculated by 

 \begin{equation}
    M_\xi=\xi + 5log\varpi + 5 - A_\xi
	\label{eq:Mzeta}
\end{equation}
where $\xi$ and $A_\xi$ are apparent magnitude and interstellar extinction in which the symbol $\xi$ indicates one of the bands at Johnson $B$, $V$, Gaia $G$, $G_{\rm BP}$, $G_{\rm RP}$ or $TESS$, while $\varpi$ is the parallax in parsec units. The uncertainties of the multiband absolute magnitudes are calculated by 

\begin{equation}
    \Delta M_\xi=\sqrt{(\Delta m_\xi)^2+(5 log e \frac{\sigma_\varpi}{\varpi})^2 + (\Delta A_\xi)^2}
	\label{eq:Muncertain}
\end{equation}
where the first term in the square root represents the uncertainty contribution of the apparent magnitude, while the second and third terms are uncertainty contributions of the parallax and interstellar extinction. Computed multiband absolute magnitudes and associated uncertainties are listed in Table \ref{tablehostBC} together with multiband $BC$ values from the 
$BC-T_{\rm eff}$ relations in Table \ref{tab:BCpartable}.

Next, the absolute bolometric magnitudes from multiband photometry are calculated by
\begin{equation}
     M_{\rm Bol}(\xi)=M_\xi + BC_\xi
	\label{eq:Mbolzeta}
\end{equation}
and listed in Table \ref{tablehostbol} together with their propagated errors from 
\begin{equation}
    \Delta M_{\rm Bol}(\xi)=\sqrt{(\Delta M_\xi)^2 + (\Delta BC_\xi)^2}
	\label{eq:DeltaMbolzeta}
\end{equation}
where $\Delta BC_\xi$ are RMS values in Table \ref{tab:BCpartable}, while $\Delta M_\xi$ are from Table \ref{tablehostBC}. Notice that the propagated errors are slightly bigger than the RMS values in Table \ref{tab:BCpartable}. This must be the result of the high accuracy at the apparent magnitudes and parallaxes of the host stars selected for this study.

Next, multiband absolute bolometric magnitudes are combined by a weighted mean using 

\begin{equation}
M_{\rm Bol} = \frac{\sum_{i=1}^{N} w_i M_{{\rm Bol},i} }{\sum_{i=1}^{N} w_i}
\label{eq:weighted}
\end{equation}
where $M_{bol,i} = M_i + BC_i$, provided with $i=$ $B$, $V$, $G$, $G_{\rm BP}$, $G_{\rm RP}$ and $TESS$ passbands. N is a number between 2 (if $M_{\rm Bol}$ is predicted from two bands only) and 6 (if $M_{\rm Bol}$ is predicted from all the passbands) while $w_i$ is the weight for each passband which is  $w_i=\frac{1}{\sigma_i^2}$, where $\sigma_i$ is the uncertainty of $M_{bol,i}$.

The weighted mean and its standard error are given in columns 15 and 16 in Table \ref{tablehostbol}.  At last, the standard $L$ of the host stars is calculated from the weighted mean in SI units according to Eq.\ref{eq:mbol1}. Then the standard $L$ of SI units are transformed into solar units. Then both $L$ by SI and solar units are recorded at columns 17 and 18 in Table \ref{tablehostbol}. The equation by \cite{Eker2021b}

\begin{equation}
\frac{\Delta L}{L}=\frac{\Delta M_{\rm Bol}}{2.5 log e}=0.921 \Delta M_{\rm Bol}
	\label{eq:DeltaL}
\end{equation}
is used for propagating the standard error of the weighted mean to estimate the uncertainty of  the standard $L$ predicted from the six-band photometry. It is listed in column 19 in Table \ref{tablehostbol}.

\begin{table*}
\centering
\caption{Absolute magnitudes and bolometric corrections at Johnson $B$, $V$, Gaia $G$, $G_{\rm BP}$, $G_{\rm RP}$ and TESS passbands.  The full table is available online.}
\resizebox{\textwidth}{!}
{\begin{tabular}{cccccccccccccccccccc}
\hline
Order	&	Hostname	&	$M_{\rm B}$	&	err	&	$M_{\rm V}$	&	err	&	$M_{\rm G}$	&	err	&	$M_{\rm GBP}$	&	err	&	$M_{\rm GRP}$	&	err	&	$M_{\rm TESS}$	&	err	&	$BC_{\rm B}*$	&	$BC_{\rm V}*$	&	$BC_{\rm G*}$	&	$BC_{\rm GBP}*$	&	$BC_{\rm GRP}*$	&	$BC_{\rm TESS}*$	\\
	&		&	[mag]	&	[mag]	&	[mag]	&	[mag]	&	[mag]	&	[mag]	&	[mag]	&	[mag]	&	[mag]	&	[mag]	&	[mag]	&	[mag]	&	[mag]	&	[mag]	&	[mag]	&	[mag]	&	[mag]	&	[mag]	\\
\hline
1	&	16 Cyg B	&	5.252	&	0.083	&	4.591	&	0.034	&	4.449	&	0.029	&	4.776	&	0.035	&	3.950	&	0.019	&	4.004	&	0.020	&	-0.609	&	0.066	&	0.106	&	-0.138	&	0.572	&	0.526	\\
2	&	24 Sex	&	2.952	&	0.044	&	2.036	&	0.038	&	1.810	&	0.029	&	2.251	&	0.035	&	1.202	&	0.020	&	1.259	&	0.021	&	-0.874	&	-0.046	&	0.078	&	-0.293	&	0.663	&	0.617	\\
3	&	2MASSJ08421149+1916373	&	6.578	&	0.041	&	5.719	&	0.086	&	5.496	&	0.030	&	5.905	&	0.036	&	4.928	&	0.021	&	4.986	&	0.022	&	-0.778	&	0.000	&	0.093	&	-0.233	&	0.639	&	0.592	\\
4	&	55 Cnc	&	6.316	&	0.044	&	5.451	&	0.038	&	5.233	&	0.029	&	5.656	&	0.035	&	4.647	&	0.019	&	4.706	&	0.020	&	-0.770	&	0.004	&	0.094	&	-0.228	&	0.636	&	0.589	\\
5	&	61 Vir	&	5.784	&	0.041	&	5.084	&	0.031	&	4.877	&	0.029	&	5.249	&	0.035	&	4.350	&	0.020	&	4.429	&	0.021	&	-0.668	&	0.046	&	0.104	&	-0.169	&	0.599	&	0.552	\\
…	&	…	&	…	&	…	&	…	&	…	&	…	&	…	&	…	&	…	&	…	&	…	&	…	&	…	&	…	&	…	&	…	&	…	&	…	&	…	\\
337	&	WASP-187	&	2.811	&	0.093	&	2.330	&	0.034	&	2.189	&	0.032	&	2.465	&	0.038	&	1.762	&	0.024	&	1.814	&	0.026	&	-0.499	&	0.091	&	0.098	&	-0.090	&	0.502	&	0.457	\\
338	&	WASP-189	&	1.810	&	0.048	&	1.619	&	0.038	&	1.593	&	0.030	&	1.680	&	0.036	&	1.409	&	0.020	&	1.438	&	0.021	&	-0.302	&	-0.019	&	-0.066	&	-0.108	&	0.125	&	0.088	\\
339	&	XO-2 N	&	6.064	&	0.166	&	5.363	&	0.032	&	5.088	&	0.030	&	5.506	&	0.035	&	4.509	&	0.020	&	4.564	&	0.021	&	-0.764	&	0.007	&	0.095	&	-0.224	&	0.634	&	0.588	\\
340	&	XO-7	&	4.123	&	0.096	&	3.485	&	0.031	&	3.442	&	0.030	&	3.705	&	0.036	&	3.044	&	0.020	&	3.093	&	0.021	&	-0.477	&	0.093	&	0.094	&	-0.082	&	0.483	&	0.439	\\
341	&	YSES 2	&	6.374	&	0.156	&	5.541	&	0.032	&	5.203	&	0.029	&	5.676	&	0.036	&	4.576	&	0.021	&	4.645	&	0.020	&	-1.074	&	-0.161	&	0.034	&	-0.431	&	0.695	&	0.651	\\
\hline
\multicolumn{20}{l}{* Uncertainties: 0$^m$.1363 ($BC_{\rm B}$), 0$^m$.1201 ($BC_{\rm V}$), 0$^m$.1107 ($BC_{\rm G}$), 0$^m$.1266 ($BC_{\rm GBP}$), 0$^m$.1092 ($BC_{\rm GRP}$) and 0$^m$.1110 ($BC_{\rm TESS}$).}
\end{tabular}}
\label{tablehostBC}
\end{table*}

\subsection{Calculating standard \texorpdfstring{$R$}{R} from standard \texorpdfstring{$L$}{L}}

Having accurately determined standard $L$ by the multiband photometry in this study, we have tried predicting the standard $R$ of a star in order to compare it to the published $R$ using

\begin{equation}
\frac{R}{R_\odot}=\sqrt{\frac{L/L_\odot}{(T_{\rm eff}/T_\odot)^4}}
	\label{eq:RoverRsun}
\end{equation}
where $L$ and $T_{\rm eff}$ are the recovered luminosity and published effective temperature of a star while $L_\odot$ is $3.8275\times10^{26}$ W and $T_\odot$=5772 K are the nominal luminosity and effective temperature of the Sun. Uncertainty of the predicted radius is obtained by 
\begin{equation}
\frac{\Delta R}{R}=\sqrt{(\frac{\Delta L}{2L})^2+(2\frac{\Delta T}{T})^2}.
	\label{eq:DeltaRoverRsun}
\end{equation}
The predicted standard $R$ and its uncertainty are given in columns 20 and 21 of Table \ref{tablehostbol}.

\begin{table*}
\centering
\caption{Bolometric absolute magnitudes from the bands of Johnson $B$, $V$, Gaia $G$, $G_{\rm BP}$, $G_{\rm RP}$, TESS passband, weighted mean, recovered luminosity and radii of the host stars. The full table is available online.}
\resizebox{\textwidth}{!}
{\begin{tabular}{cccccccccccccccccccccc}
\hline
Order	&	Hostname	&	$M_{\rm bol}(B)$	&	err	&	$M_{\rm bol}(V)$	&	err	&	$M_{\rm Bol}(G)$	&	err	&	$M_{\rm Bol}(G_{\rm BP})$	&	err	&	$M_{\rm Bol}(G_{\rm RP})$	&	err	&	$M_{\rm Bol}(TESS)$	&	err	&	Mean $M_{\rm Bol}$	&	Std err	&	$L/L_\odot$	&	$L$(SI)	&	$\frac{\Delta L}{L}$	&	$R$	&	err	\\
	&		&	[mag]	&	[mag]	&	[mag]	&	[mag]	&	[mag]	&	[mag]	&	[mag]	&	[mag]	&	[mag]	&	[mag]	&	[mag]	&	[mag]	&	[mag]	&	[mag]	&		&	$\times 10^{27}$	&	[\%]	&	[$R_\odot$]	&	[$R_\odot$]	\\
\hline
1	&	16 Cyg B	&	4.643	&	0.159	&	4.657	&	0.125	&	4.555	&	0.114	&	4.638	&	0.131	&	4.522	&	0.111	&	4.529	&	0.113	&	4.580	&	0.023	&	1.159	&	0.443	&	0.021	&	1.085	&	0.012	\\
2	&	24 Sex	&	2.078	&	0.143	&	1.990	&	0.126	&	1.888	&	0.115	&	1.958	&	0.131	&	1.866	&	0.111	&	1.876	&	0.113	&	1.930	&	0.031	&	13.309	&	5.094	&	0.028	&	4.677	&	0.105	\\
3	&	2MASSJ08421149+1916373	&	5.800	&	0.142	&	5.719	&	0.148	&	5.589	&	0.115	&	5.672	&	0.132	&	5.566	&	0.111	&	5.577	&	0.113	&	5.636	&	0.035	&	0.438	&	0.168	&	0.032	&	0.785	&	0.015	\\
4	&	55 Cnc	&	5.546	&	0.143	&	5.455	&	0.126	&	5.327	&	0.114	&	5.428	&	0.131	&	5.284	&	0.111	&	5.296	&	0.113	&	5.373	&	0.039	&	0.558	&	0.214	&	0.036	&	0.880	&	0.031	\\
5	&	61 Vir	&	5.117	&	0.142	&	5.131	&	0.124	&	4.981	&	0.114	&	5.080	&	0.131	&	4.949	&	0.111	&	4.982	&	0.113	&	5.029	&	0.029	&	0.766	&	0.293	&	0.027	&	0.938	&	0.017	\\
…	&	…	&	…	&	…	&	…	&	…	&	…	&	…	&	…	&	…	&	…	&	…	&	…	&	…	&	…	&	…	&	…	&	…	&	…	&		&	…	\\
337	&	WASP-187	&	2.312	&	0.165	&	2.421	&	0.125	&	2.287	&	0.115	&	2.375	&	0.132	&	2.264	&	0.112	&	2.271	&	0.114	&	2.316	&	0.024	&	9.322	&	3.568	&	0.022	&	2.689	&	0.083	\\
338	&	WASP-189	&	1.507	&	0.144	&	1.601	&	0.126	&	1.527	&	0.115	&	1.572	&	0.132	&	1.535	&	0.111	&	1.525	&	0.113	&	1.544	&	0.013	&	18.993	&	7.269	&	0.012	&	2.269	&	0.047	\\
339	&	XO-2 N	&	5.300	&	0.215	&	5.370	&	0.124	&	5.183	&	0.115	&	5.282	&	0.131	&	5.143	&	0.111	&	5.151	&	0.113	&	5.221	&	0.034	&	0.642	&	0.246	&	0.032	&	0.939	&	0.025	\\
340	&	XO-7	&	3.646	&	0.167	&	3.578	&	0.124	&	3.536	&	0.115	&	3.623	&	0.132	&	3.528	&	0.111	&	3.533	&	0.113	&	3.563	&	0.019	&	2.956	&	1.131	&	0.017	&	1.466	&	0.049	\\
341	&	YSES 2	&	5.300	&	0.207	&	5.381	&	0.124	&	5.237	&	0.115	&	5.245	&	0.132	&	5.271	&	0.111	&	5.296	&	0.113	&	5.286	&	0.019	&	0.605	&	0.232	&	0.018	&	1.149	&	0.022	\\
\hline
\end{tabular}}
\label{tablehostbol}
\end{table*}

\begin{figure}
\begin{subfigure}[b]{.38\textwidth}
  \centering
\includegraphics[width=\textwidth]{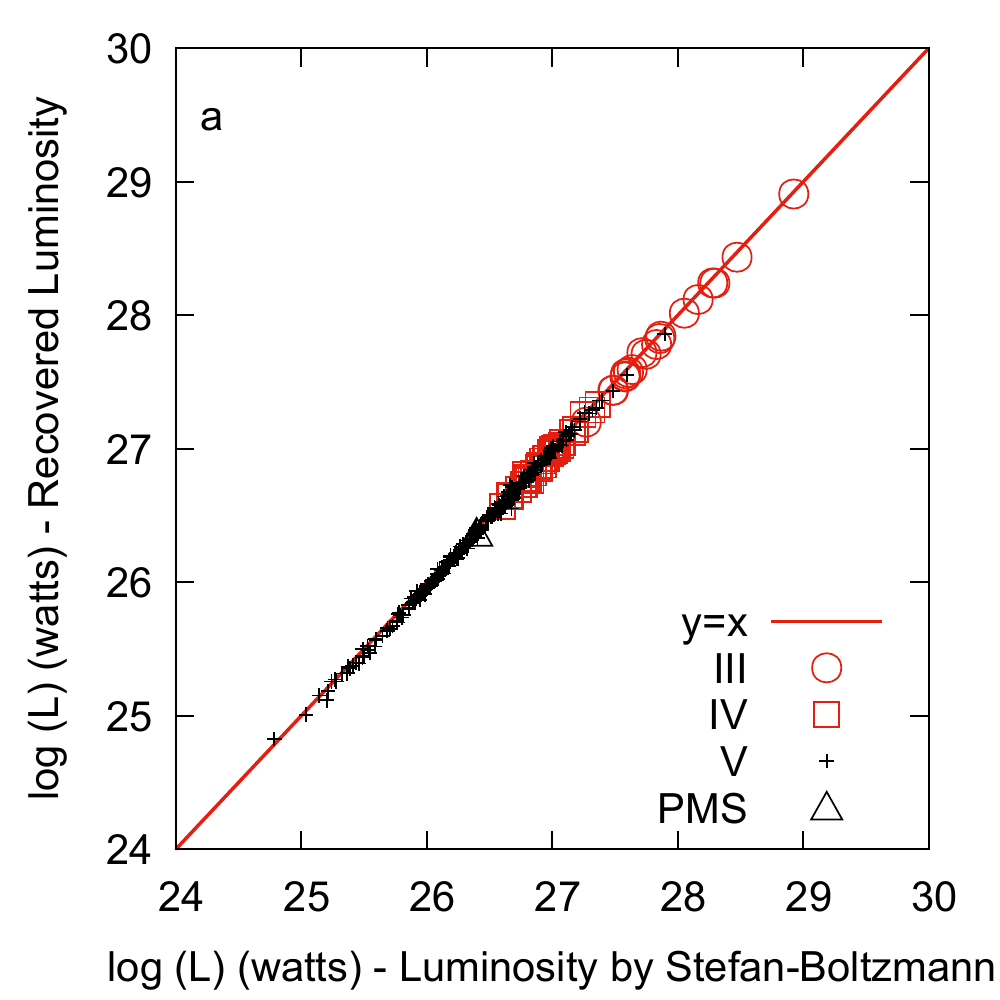}
\end{subfigure}
\hfill
\begin{subfigure}[b]{.38\textwidth}
  \centering \includegraphics[width=\textwidth]{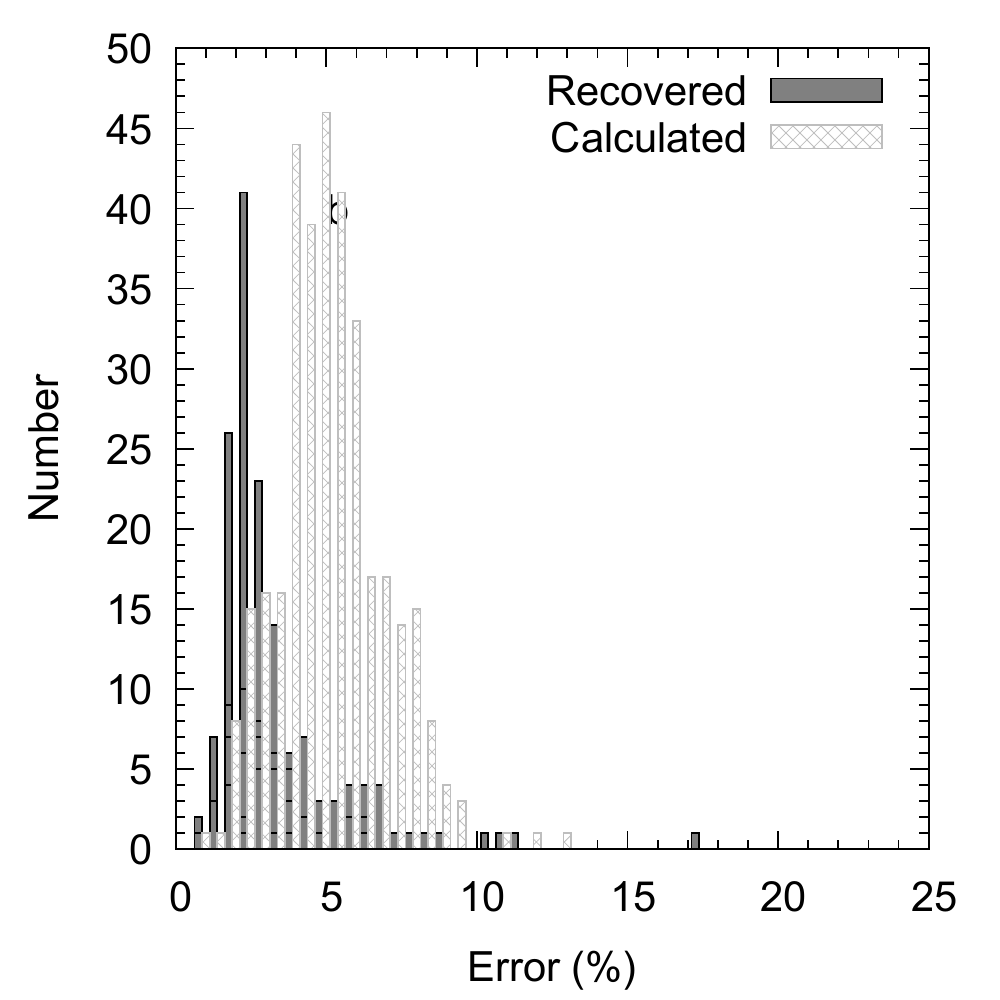}
\end{subfigure}
\caption{a) Comparing recovered $L$ (from photometry) and calculated $L$  (from $R$ and $T_{\rm eff}$) of the sample stars. b) Histogram distribution of the uncertainties associated with recovered (dark) $L$ is compared to the histogram distribution of the uncertainties associated with calculated (grey) $L$.}
\label{fig:luminosity_errors}
\end{figure}

\begin{figure}
\begin{subfigure}[b]{.4\textwidth}
  \centering
\includegraphics[width=\textwidth]{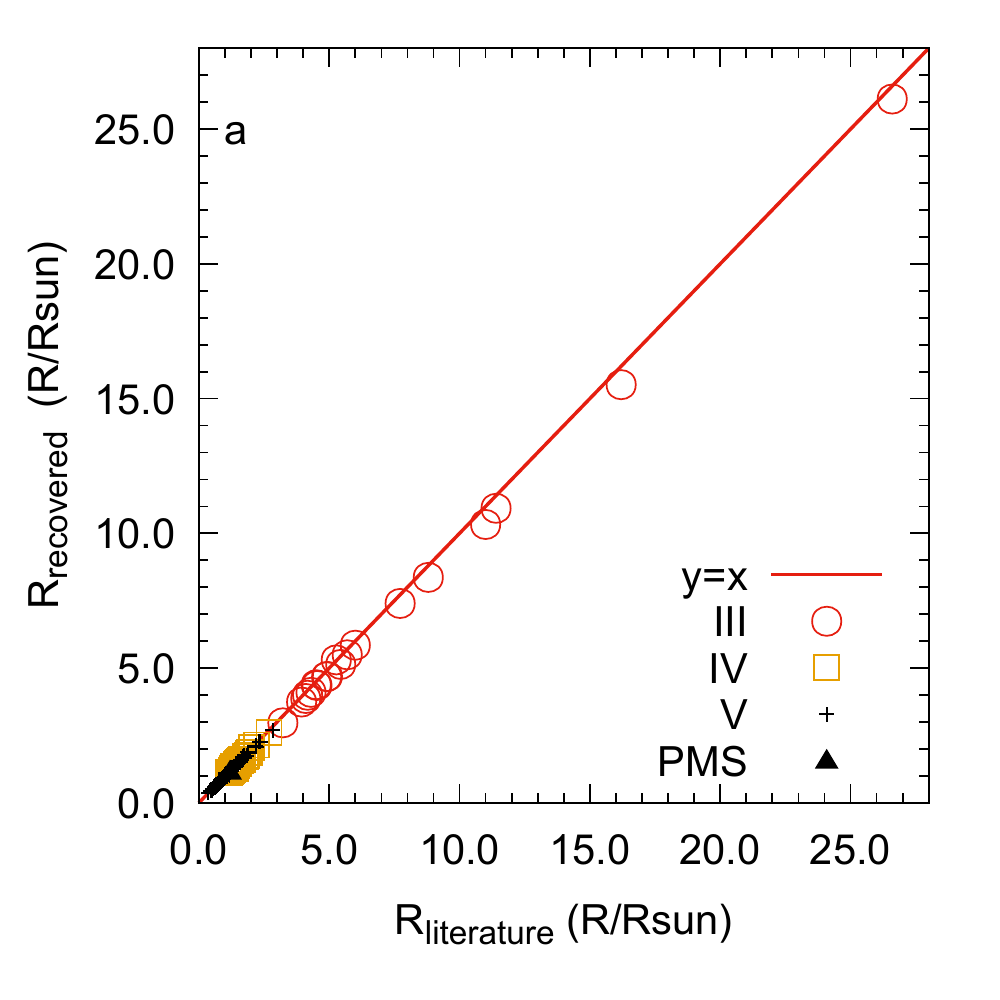}
\end{subfigure}
\hfill
\begin{subfigure}[b]{.4\textwidth}
  \centering \includegraphics[width=\textwidth]{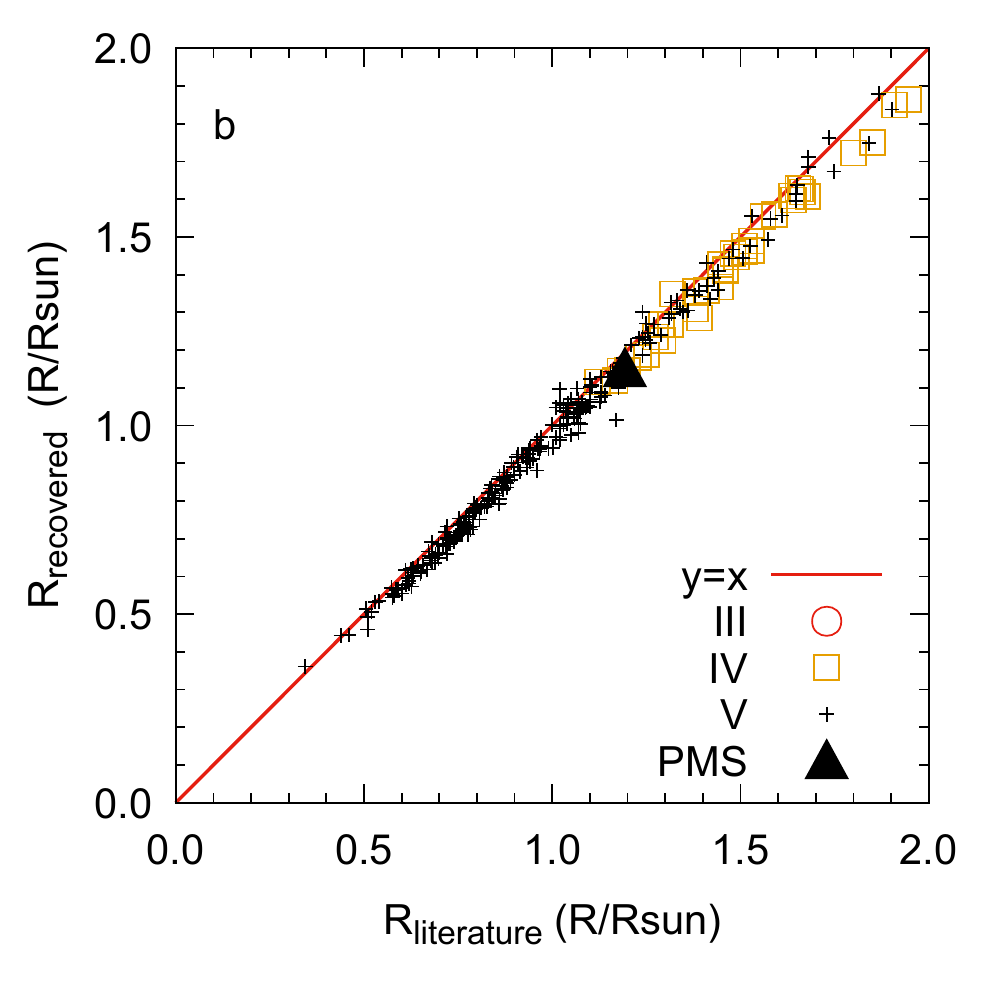}
\end{subfigure}
\hfill
\begin{subfigure}[b]{.4\textwidth}
  \centering \includegraphics[width=\textwidth]{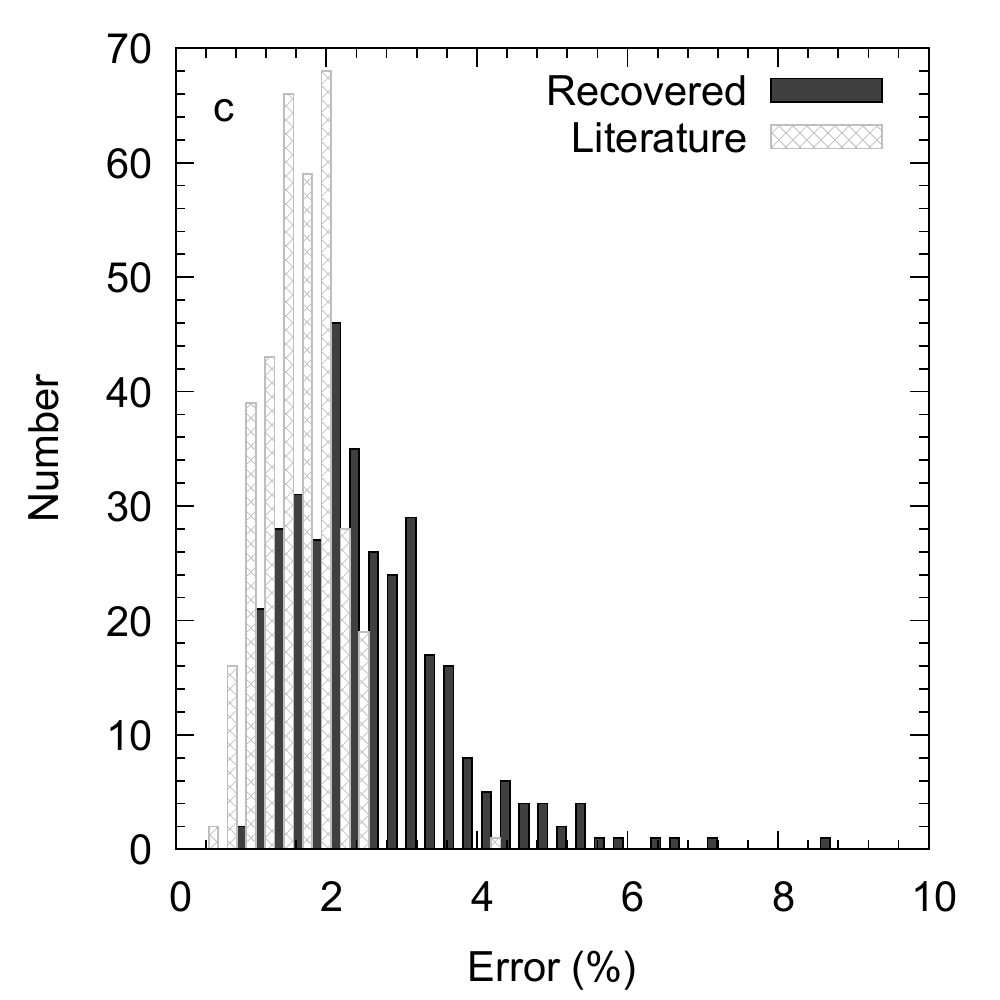}
\end{subfigure}
\caption{Comparing recovered and published R of giants (a), subgiants (b), main-sequence stars and a PMS star. Error histograms of recovered and published $R$ (c).}
\label{fig:radii}
\end{figure}

\section{Discussions}

It is known that $BC$ of a star does not only vary by its spectral type and/or $T_{\rm eff}$ but also varies by its luminosity class (log $g$) \citep{Johnson1966, Cox2000} and metallicity [m/H] \citep{Girardi2002, Girardi2008, Masana2006, Pedersen2020} and even by its rotation speed \citep{Chen2019}. However, the empirical $BC$ values and the $BC - T_{\rm eff}$ relations of \cite{Flower1996}, which were rectified later by \cite{Torres2010a}, was claimed unique for all luminosity classes; main-sequence (V), subgiants (IV), giants (III), supergiants (II) and bright supergiants (I). Unfortunately, being deduced from the 335 stars of mixed luminosity collected from the literature, the data was not sufficient to calibrate various $BC-T_{\rm eff}$ relations of different luminosity classes. 

Therefore, the original and the rectified relations were claimed to be valid at all luminosity classes in the temperature range from 2900 to 52000 K which were expressed by three different polynomials each covering a temperature range as log $T_{\rm eff}$ $<$ 3.70, 3.70 $<$ log $T_{\rm eff}$ $<$ 3.90 and log $T_{\rm eff}$ $>$ 3.90 by a note “The $T_{\rm eff}$:$BC$ relation for stars with temperatures less than $\sim$5000 K, however, is uncertain”. \cite{Torres2010a} quoted “Many authors find this source convenient because, in addition to the table, it presents simple polynomial fits for $BC_{\rm V}$ valid over a wide range of temperatures.”, therefore, the claim of \cite{Flower1996} is critical to be tested for the sake of $BC$ users in various fields of astrophysics.

\cite{Flower1996} was the only study giving $BC_{\rm V}$ of a star as a function of its $T_{\rm eff}$ while other sources of $BC$ were in tabulated forms until \cite{Eker2020} who calibrated a single main-sequence $BC_{\rm V}-T_{\rm eff}$ relation valid in the temperature range 3100 - 36000 K using the published parameters of 290 DDEB systems having Gaia DR2 trigonometric parallaxes. Non-main-sequence empirical $BC-T_{\rm eff}$ relations are still not calibrated to confirm theoretically claimed differences of a $BC$ at various luminosity classes which appear as the tabulated tables generated  theoretically using model atmospheres.

 The situation has not yet changed, thus, one way of testing the claim of \cite{Flower1996} is to apply most recently calibrated main-sequence empirical multi-band $BC -T_{\rm eff}$ relations of \cite{Bakis2022} and $BC_{\rm TESS}-T_{\rm eff}$ relation of this study on the stars with most accurate $R$ and $T_{\rm eff}$ with mixed luminosity classes and to see how successful their $L$ would be recovered from their multiband apparent magnitudes (Johnson $B$, $V$, Gaia $G$, $G_{\rm BP}$, $G_{\rm RP}$ and $TESS$), $BC$ values (Table \ref{tab:BCpartable}), parallaxes (Table \ref{tablehostphysics}) and interstellar extinctions (Table \ref{tablehostmags}). 
 
 This study gave us this opportunity to compare the recovered $L$ from multiband photometry and the computed $L$ from published $R$ and $T_{\rm eff}$ by the Stefan Boltzmann law.

\subsection{Comparing \texorpdfstring{$L$}{L} from multiband photometry and \texorpdfstring{$L$}{L} from \texorpdfstring{$R$}{R} and \texorpdfstring{$T_{\rm eff}$}{Teff}}

Figure \ref{fig:luminosity_errors}a compares the standard $L$ of the host stars (Table \ref{tablehostbol}, column 18) recovered from their six band (Johnson $B$, $V$, Gaia $G$, $G_{\rm BP}$, $G_{\rm RP}$ and $TESS$) photometry and $L$ (Table \ref{tablehostphysics}, column 15) of the same sample calculated directly from their published $R$ and $T_{\rm eff}$ using the Stefan-Boltzmann law. Regardless to the luminosity classes of the host stars, an outstanding agreement between the predicted and the computed $L$ values is clear.

Accuracy of the recovered $L$ is indicated in Figure \ref{fig:luminosity_errors}b, where histogram error distributions of the recovered $L$ of host stars are compared to the histogram error distributions of the computed $L$.  It is obviously seen in Figure \ref{fig:luminosity_errors}b that the indirect method of obtaining $L$, which requires a pre-computed $BC$, produced twice more accurate standard $L$ than the direct method because the peak of error distribution of the predicted $L$ is 2 per cent while the peak of error distribution of the computed $L$ is 4 per cent. This result, however, is the result of combining independent photometric $M_{\rm Bol}$ values (Table \ref{tablehostbol}) by a weighted average, where the standard error of the weighted mean is propagated as the relative uncertainty ($\Delta L/L$) of the recovered $L$. If single band photometry is used to obtain a single $M_{\rm Bol}$ from $M_\xi$ + $BC_\xi$, where $\xi$ is one of the bands of Johnson $B$, $V$, Gaia $G$, $G_{\rm BP}$, $G_{\rm RP}$ and $TESS$, the accuracy of the recovered $L$ would have been worse; worse than the errors of computed $L$ which has a peak at 4 per cent. 

Because of very accurate apparent magnitudes, parallaxes and a few per cent accuracies in interstellar extinctions (Table \ref{tablehostmags}), the  errors of the absolute magnitudes in Table \ref{tablehostBC} contributed very little with respect to the error contributions of $BC$s that the errors of $M_{\rm Bol}(\xi)$ in Table \ref{tablehostbol} are just slightly bigger than the corresponding RMS values in Table \ref{tab:BCpartable}. According to Table \ref{tablehostbol} the most accurate $M_{\rm Bol}(\xi)$ would have a 0.111 mag uncertainty, which corresponds to 10 per cent error in the recovered $L$ ($\Delta L/L$) according to Eq.\ref{eq:DeltaL} if single channel photometry is used. If a non-standard $BC$s were used, an additional 10 per cent error further reducing this accuracy  would have been inevitable \citep[see][]{Torres2010a, Eker2021a, Eker2021b, Eker2022}, which was the case in earlier applications with various tabulated $BC$ values. 

Figure \ref{fig:luminosity_errors} indicates that a main-sequence $BC$ or $BC - T_{\rm eff}$ relation could be used in predicting  $L$ of giants, subgiants and even a PMS also without ruling out the fact that $BC$ of a star also depends on its luminosity class (log $g$), metal abundance and even its rotational speed as displayed by many tabulated $BC$s produced theoretically from model atmospheres. This fact becomes more noticeable in the comparison of the predicted $R$ and published $R$ of the host stars in the next subsection.

\subsection{Comparing predicted \texorpdfstring{$R$}{R} to published \texorpdfstring{$R$}{R}}
Figure \ref{fig:radii}ab compares the predicted $R$  (Table \ref{tablehostbol}, column 20) from published $T_{\rm eff}$ and recovered $L$  to the published $R$ (Table \ref{tablehostphysics}, column 6) of the sample host stars. A sufficiently good agreement between the predicted and published $R$ appears supporting \cite{Flower1996} but, on the other hand, a small but clear offset of giants (Fig \ref{fig:radii}a) and subgiants (Fig \ref{fig:radii}b) from the diagonal most probably caused by small difference of $BC$ values among different luminosity classes actually disapproves him. The Small difference appears to be negligible so that a user could be satisfied by the recovered R of a giant star using a main-sequence $BC - T_{\rm eff}$ relation, but this does not mean $BC - T_{\rm eff}$ relations of different luminosity classes are the same.    

Accuracy of the predicted $R$ of the whole sample is displayed in Figure \ref{fig:radii}c, where error histograms of the predicted and published  $R$ of the host stars are compared. It is obvious in Figure \ref{fig:radii}c that the peaks of both distributions are about the same at 2 per cent, however, the error distribution of the predicted $R$ has rather a gradually decreasing tail reaching out to 6 per cent while the errors of published $R$  has a sharply ending tail at 2.5 per cent due to the selection criteria of the host stars for this study. More accurate sides of the distributions are also different so that almost half of the most accurate $R$s of the host stars appear to be moved towards the less accurate side because of the less accurate tail of the $L$ errors (see Fig.\ref{fig:histLRT}).

Such an error distribution of the predicted $R$ is deducible from Eq.\ref{eq:DeltaRoverRsun}. Noticing that the peak of the error distribution of the predicted $L$ is at 2 per cent (see Fig. \ref{fig:luminosity_errors}b) and the peak of the error distribution of the published $T_{\rm eff}$ (See Fig \ref{fig:histLRT}a) is about 1 per cent. Plugging these values into Eq.\ref{eq:DeltaRoverRsun}, one obtains the typical error of predicted $R$ is $\sim2$ per cent ($\sqrt{5}$). It is obvios that the errors of published $T_{\rm eff}$ dominate.  That is, using multiband $BC-T_{\rm eff}$ relations in this study, it is now possible to obtain $R$ of a single star with a relative error twice the relative error of $T_{\rm eff}$.

\subsection{How a main-sequence \texorpdfstring{$BC-T_{\rm eff}$}{BC-Teff} relation works at all luminosity classes?}

Main-sequence $BC - T_{\rm eff}$ relations from Table \ref{tab:BCpartable} are used in predicting $L$ and $R$ of the host stars. Figures \ref{fig:luminosity_errors} \& \ref{fig:radii} indicate predictions are successful within the error limits regardless of the luminosity classes (V, IV, III and a PMS) of the host stars. Therefore, one may think the claim of \cite{Flower1996} "All luminosity classes appear to follow a unique $BC$ - $Teff$ relation" would be true.    
Consequently, a reader would have a question “How do main-sequence $BC - T_{\rm eff}$ relations work in predicting the $L$ and $R$ of a star, which could be a main sequence star, or a subgiant or a giant and even a pre-main-sequence star?”.

This is because the information about the luminosity class of a star is contained mostly in its apparent magnitude rather than its $BC$. Therefore, because the contributions of $BC$s are secondary or negligible in the predictions of $M_{\rm Bol}(\xi)$, which are $M_\xi + BC_\xi$, one can apparently obtain standard $L$ and $R$ of a single star, which may belong to any luminosity class V, IV, III, or PMS, from its multiband photometry within the error limits set by the propagation of observational random errors and RMS deviations of the $BC-T_{\rm eff}$ relations (Table \ref{tab:BCpartable}) if the star has a reliable $T_{\rm eff}$ and parallax.

\section{Conclusions}
The following main-sequence $BC_{\rm TESS} - T_{\rm eff}$ relation, where $X$ = log $T_{\rm eff}$.

\begin{equation}
BC_{\rm TESS} = -318.533 + 232.298X -55.2916X^2 + 4.27613X^3
	\label{eq:BCTESS}
\end{equation}
is calibrated using the published $R$ and $T_{\rm eff}$ of 390 main-sequence stars which are the components of 202 DDEB with TESS apparent magnitudes selected from 209 DDEB. The other five band (Johnson $B$, $V$, Gaia $G$, $G_{\rm BP}$ and $G_{\rm RP}$) $BC - T_{\rm eff}$ relations (Table \ref{tab:BCpartable}) calibrated by the same DDEB sample are taken from \cite{Bakis2022}. Being different from previously calibrated relations, which are polynomials of the fourth degree, this newly calibrated relation is found to be a third-degree polynomial fitting best to the existing data. The uncertainties of the coefficients and other statistics such as RMS deviations and correlation coefficient ($R^2$) are given in Table \ref{tab:BCpartable} together with the other statistics of the other five band-relations.

The five-band $BC - T_{\rm eff}$ relations and $BC$ coefficients were already tested by \cite{Bakis2022} by a successful recovery of $L$ of the main-sequence stars from the same DDEB sample used in the calibrations of these relations. Consequently, \cite{Bakis2022} concluded that one of the secondary methods of obtaining $L$, which requires a pre-computed $BC$, may provide more accurate $L$ than the classical method relying on the Stefan-Boltzmann law if the information provided by multiband $BC-T_{\rm eff}$ relations are combined. Briefly, various $M_{\rm Bol}$ are calculated first using the $BC$ values from the multiband $BC - T_{\rm eff}$ relation for the given $T_{\rm eff}$ of a star. Then all $M_{\rm Bol}$ values are combined to a mean $M_{\rm Bol}$. Eq.\ref{eq:mbol1} is used to obtain $L$ of the star from the mean $M_{\rm Bol}$. Including the $BC_{\rm TESS} - T_{\rm eff}$ relation produced in this study, the six-band (Johnson $B$, $V$, Gaia $G$, $G_{\rm BP}$, $G_{\rm RP}$ and $TESS$) $BC - T_{\rm eff}$ relations, which are deduced from DDEB components, were tested by recovering the $L$ and $R$ of the single host stars with the most accurate published $T_{\rm eff}$ and $R$ similarly.

Both tests (recovering $L$ and recovering $R$) are found successful to conclude that 1) Using multiband $BC - T_{\rm eff}$ relations and $T_{\rm eff}$, one can obtain $L$ of a single star more accurate than the classical direct method of obtaining $L$ relying on the Stefan-Boltzmann law.  2) Accurately predicted $L$ of a single star could be used in predicting its $R$, a critical parameter for exoplanet studies, with reasonable accuracy. The $T_{\rm eff}$  of the star is the only stellar parameter one needs to know to obtain  its  $L$ and $R$ in addition to its parallax, apparent magnitudes in Johnson $B$, $V$, Gaia $G$, $G_{\rm BP}$, $G_{\rm RP}$ and $TESS$ passband and interstellar extinctions.

Since the present sample contains stars of various luminosity classes (281 main-sequence, 40 subgiants, 19 giants and 1 PMS), we could conclude that not only $L$ and $R$ of the main-sequence but also subgiant, giant and even pre-main sequence stars could be recovered at about 2 per cent uncertainty if the input $T_{\rm eff}$ in the range from 2900 to 38000 K is accurate to have a one per cent uncertainty using the six band $BC - T_{\rm eff}$ relations in Table \ref{tab:BCpartable}, despite they were calibrated with main-sequence stars.

This result implies that the claim of \cite{Flower1996} "All luminosity classes appear to follow a unique $BC$ - $Teff$ relation" is true, however, it does not rule out the effect of $log  g$ of a star on its $BC$. This is because there is a small but clear underestimation with the recovered $R$ of the subgiant and giant hosts according to Figure \ref{fig:radii} and perhaps a similar underestimation of recovered $L$ of the same subgiant and giant hosts also exists in Figure \ref{fig:luminosity_errors}. These negligible differences may well be caused by  small differences in $BC$ values due to different luminosity classes. Thus, we can also conclude that the information about the luminosity class of a star is mostly contained in its apparent magnitude rather than its $BC$. Thus, a main sequence $BC$ or a main-sequence $BC - T_{\rm eff}$ relation could be used in predicting $L$ and $R$ of single stars in all luminosity classes.

We encourage researchers to calculate $BC$ values of a sufficient number of non-main sequence stars and calibrate empirical $BC - T_{\rm eff}$ relation also for giants, subgiants, and PMS in order to see the difference and to confirm the theoretically produced tabulated $BC$s, where the $BC$ values usually presented as functions of luminosity class (log $g$) \citep{Johnson1966, Cox2000} and metallicity [m/H] \citep{Girardi2002, Girardi2008, Masana2006, Pedersen2020} and even speed of its rotation \citep{Chen2019}.

\section*{Acknowledgements}

This work uses the VizieR catalogue access tool, CDS, Strasbourg, France; the SIMBAD database, operated at CDS, Strasbourg, France. This work presents results from the European Space Agency (ESA) space mission, Gaia. Gaia data are being processed by the Gaia Data Processing and Analysis Consortium (DPAC). Funding for the DPAC is provided by national institutions, in particular, the institutions participating in the Gaia MultiLateral Agreement (MLA). The Gaia mission website is https://www.cosmos.esa.int/gaia. The Gaia archive website is https://archives.esac.esa.int/gaia. We thank to Yasin Yal\c{c}{\i}n and Sezer Kayac{\i} for their assistance during the data collection of single host stars. Finally, we thank to Oleg Malkov who reviewed the paper with his useful comments.

\section*{Data Availability}

The data underlying this article are available in the article and in its online supplementary material.



\bibliographystyle{mnras}
\bibliography{mybib} 




\appendix




\bsp	
\label{lastpage}
\end{document}